%% file: IsingIsingStar_arxiv_Single.tex
\documentclass[prl,showpacs,reprint,amssymb,amsmath,superscriptaddress,longbibliography]{revtex4-1} %,linenumbers
\usepackage[utf8x]{inputenc}
\usepackage{graphicx}
\usepackage{xcolor}
\usepackage{colortbl}
\usepackage{grffile}		% process dots etc. in filenames for figures
\usepackage[colorlinks,bookmarks=false,citecolor=blue,linkcolor=red,urlcolor=blue]{hyperref}
\usepackage{times}
\usepackage{placeins}

\graphicspath{{figures/}}

\newcommand{\figref}[1]{Fig.~\ref{#1}}
\renewcommand{\eqref}[1]{Eq.~(\ref{#1})}

\definecolor{lightgray}{gray}{0.85}
\newcommand{\beq}{\begin{equation}}
\newcommand{\eeq}{\end{equation}}
\newcommand{\ba}{\begin{array}{ccc}}
\newcommand{\ea}{\end{array}}
\newcommand{\nn}{\nonumber \\}
\def\bea{\begin{eqnarray}}
\def\eea{\end{eqnarray}}

\begin{document}
%%%%%%%%%%%%%%%%%%%%%%%%%% TITLE ETC %%%%%%%%%%%%%%%%%%%%%%%%%%%%%%%%
\title{Universal Signatures of Quantum Critical Points from Finite-Size Torus Spectra:\\
A Window into the Operator Content of Higher-Dimensional Conformal Field Theories}
\author{Michael Schuler}
\affiliation{Institut f\"ur Theoretische Physik, Universit\"at Innsbruck, A-6020 Innsbruck, Austria}
\author{Seth Whitsitt}
\affiliation{Department of Physics, Harvard University, Cambridge, Massachusetts, 02138, USA}
\author{Louis-Paul Henry}
\affiliation{Institut f\"ur Theoretische Physik, Universit\"at Innsbruck, A-6020 Innsbruck, Austria}
\author{Subir Sachdev}
\affiliation{Department of Physics, Harvard University, Cambridge, Massachusetts, 02138, USA}
\affiliation{Perimeter Institute for Theoretical Physics, Waterloo, Ontario N2L 2Y5, Canada}
\author{Andreas M. L\"auchli}
\affiliation{Institut f\"ur Theoretische Physik, Universit\"at Innsbruck, A-6020 Innsbruck, Austria}

%%%%%%%%%%%%%%%%%%%%%%%%%% ABSTRACT %%%%%%%%%%%%%%%%%%%%%%%%%%%%%%%%
\begin{abstract}
The low-energy spectra of many body systems on a torus, of finite size $L$, are well understood in magnetically ordered and gapped
topological phases. However, the spectra at quantum critical points separating such phases are largely unexplored for 2+1D systems.
Using a combination of analytical and numerical techniques, we accurately calculate and analyse the low-energy torus spectrum at an Ising critical point which provides a universal fingerprint of the underlying quantum field theory, with the energy levels given by universal numbers times $1/L$. We highlight the implications of a neighboring topological phase on the spectrum by studying the Ising* transition (i.e. the transition between a $Z_2$ topological phase and a trivial paramagnet), in the example of the toric code in a longitudinal field, and advocate a phenomenological picture that provides qualitative insight into the operator content of the critical field theory.	
\end{abstract}
\date{\today}

%%%%%%%%%%%%%%%%%%%%%%%%%% MAKE TITLE %%%%%%%%%%%%%%%%%%%%%%%%%%%%%%%%
\maketitle

%%%%%%%%%%%%%%%%%%%%%%%%%% INTRODUCTION %%%%%%%%%%%%%%%%%%%%%%%%%%%%%%%%
\paragraph{Introduction ---}
Quantum critical points continue to attract tremendous attention in condensed
matter, statistical mechanics and quantum field theory alike. Recent
highlights include the discovery of quantum critical points which lie beyond
the Ginzburg-Landau paradigm~\cite{Senthil2004,Sandvik2007}, the striking success of the conformal bootstrap
program for Wilson-Fisher fixed points~\cite{El-Showk2012,*El-showk2014}, and the intimate connection between
entanglement quantities and universal data of the critical quantum field theory~\cite{Holzhey1994,Calabrese2004,Laeuchli2013,Kallin2013,Bueno2015}.

\begin{figure}
 \centerline{\includegraphics[height=5cm]{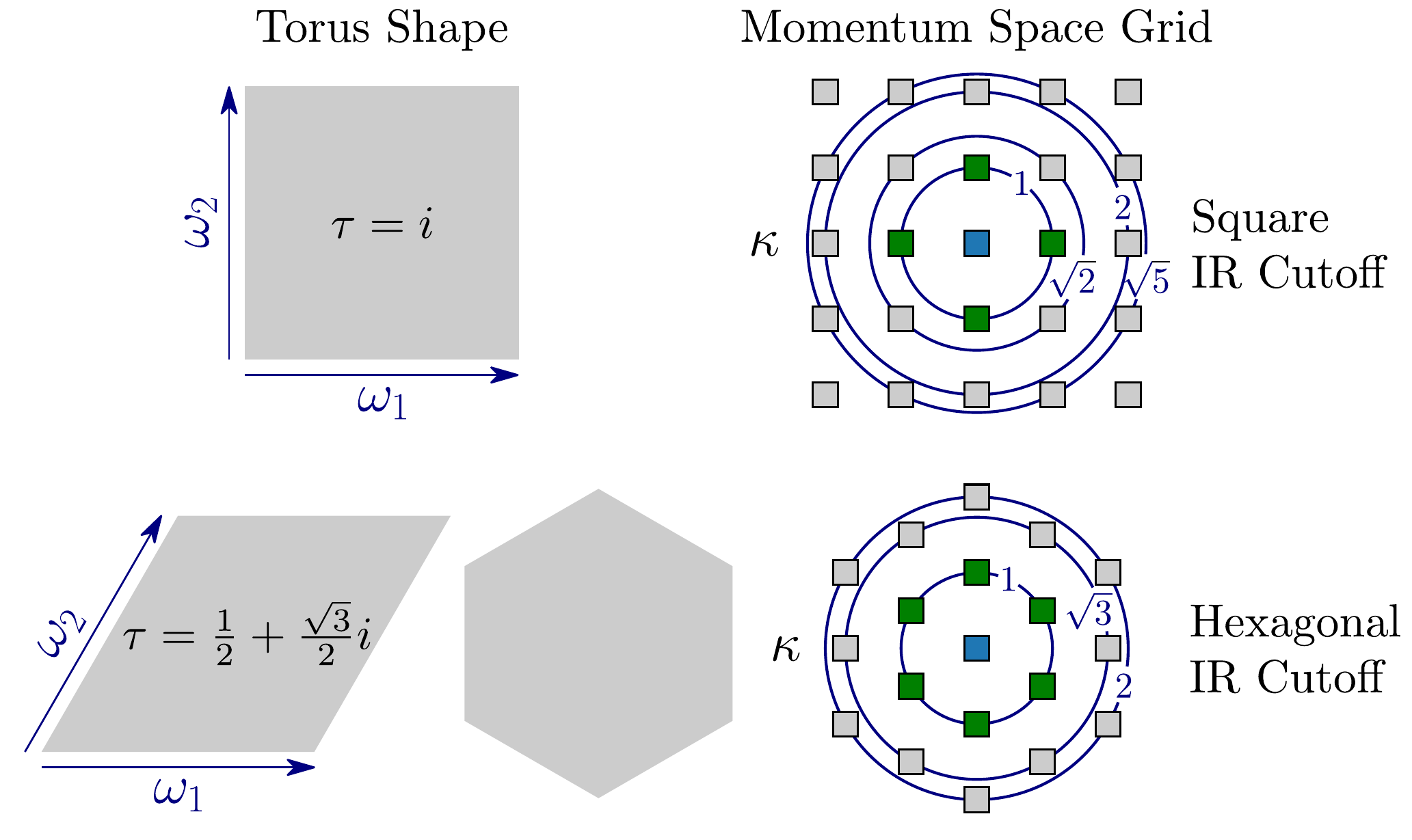}}
 \caption{\label{fig:TFI_geometry} The two torus geometries with 4-fold and 6-fold rotation symmetry and their momentum-space 
 grid in the vicinity of the $\Gamma=(0,0)$ point. In the center of the lower row we display the Wigner-Seitz cell of the torus, highlighting the 6-fold
 symmetry. The momentum space variable $\kappa$ is defined as $\kappa = \frac{L}{2\pi}|\mathbf{k}|\tau_2$ with $\tau=\tau_1 + i \tau_2$, $L=|\omega_1|=|\omega_2|$ and $\mathbf{k}$ a momentum of the finite-size cluster.}
\end{figure}

A surprisingly little explored aspect in this regard is the finite (spatial) volume spectrum on numerically easily accessible 
geometries, such as the Hamiltonian spectrum on a 2D spatial torus at the quantum critical 
point~\footnote{In a corresponding classical statistical mechanics 
language, we are discussing the spectrum of the logarithm of the transfer matrix in the limit of an infinitely long 
square (or hexagonal) rod~(c.f.~left part of Fig.~\ref{fig:TFI_geometry}). The
transfer matrix acts along the infinite rod direction.}.
In the realm of 1+1D conformal critical points 
there exists a celebrated mapping between the spectrum of scaling dimensions of the field theory in $\mathbb{R}^2$ 
and the Hamiltonian spectrum on a circle (space-time cylinder: $S^1\times\mathbb{R}$)~\cite{Cardy1984}. This result is routinely used to
perform accurate numerical spectroscopy of conformal critical points using a variety of numerical
methods~\cite{Feiguin2007,Suwa2015}. In higher dimensions the situation is less favorable: 
Cardy has shown~\cite{Cardy1985} that the corresponding conformal map can be generalized
to a map between $\mathbb{R}^d$ and $S^{d-1}\times \mathbb{R}$. While numerical simulations
in this so-called {\em radial quantization} geometry have been attempted at several occasions~\cite{Alcaraz1987b,Weigel2000c,Deng2002,Brower2013c,Brower2016}, 
this numerical approach remains very challenging due to the curved geometry, which is inherently
difficult to regularize in numerical simulations.

Although low-energy spectra on different toroidal configurations
have been discussed in the context of some specific field theories (in
Euclidean spacetime)~\cite{Henningson2007,
Henningson2011,Banerjee2013,Perez2013,Shaghoulian2016}, our understanding of
critical energy spectra is rather limited beyond free
theories~\cite{Henkel1986,*Henkel1987,Cardy1987,Hamer1999,*Hamer1986,*Hamer2000,
Nishiyama2008,Dusuel2010}. This is due to the absence of
a known relation between the scaling dimensions of the field theory and the
torus energy spectra.

In this Letter we present a combined numerical and analytical study of the Hamiltonian torus
energy spectrum of the 3D Ising conformal field theory (CFT), and show that it is accessible with finite lattice studies and proper finite-size scaling. 
Torus energy spectra provide a universal fingerprint of the quantum field theory governing the critical point and
depend only on the universality class of the transition and on the shape and boundary conditions of the torus, which acts as an infrared (IR) cutoff
(but not on the lattice discretisation, i.e.~the ultraviolet cutoff). We will explicitly demonstrate this here for the Ising CFT.
This approach will also be valuable as a new numerical tool to investigate and discriminate quantum critical points. 

We provide a quantitative analysis of many low-lying energy levels of the standard $\mathbb{Z}_2$-symmetry 
breaking phase transition in the 3D Ising universality class. We also advocate a phenomenological picture 
that provides qualitative insight into the operator content of the critical point. As an application we reveal that the torus 
energy spectrum of the confinement transition between the $\mathbb{Z}_2$ topological ordered phase and 
the trivial (confined) phase of the Toric code (TC) in a longitudinal magnetic field can be understood as a 
specific combination of a subset of the fields and several boundary conditions of the standard 3D Ising 
universality class. Since the operator content of the partition function at criticality obviously differs from the 
standard 3D Ising universality class we term this transition a 3D {\em Ising*} transition~\cite{Jalabert1991,SSMV99,Senthil2000}.

%%%%%%%%%%%%%%%%%%%%%%%%%%%%
%%%%%%%%%%%%%%%%%%%%%%%%%%%%

%%%%%%%%%%%%%%%%%%%%%%%%%%%%
\begin{figure*}
 \centerline{\includegraphics[height=4.8cm]{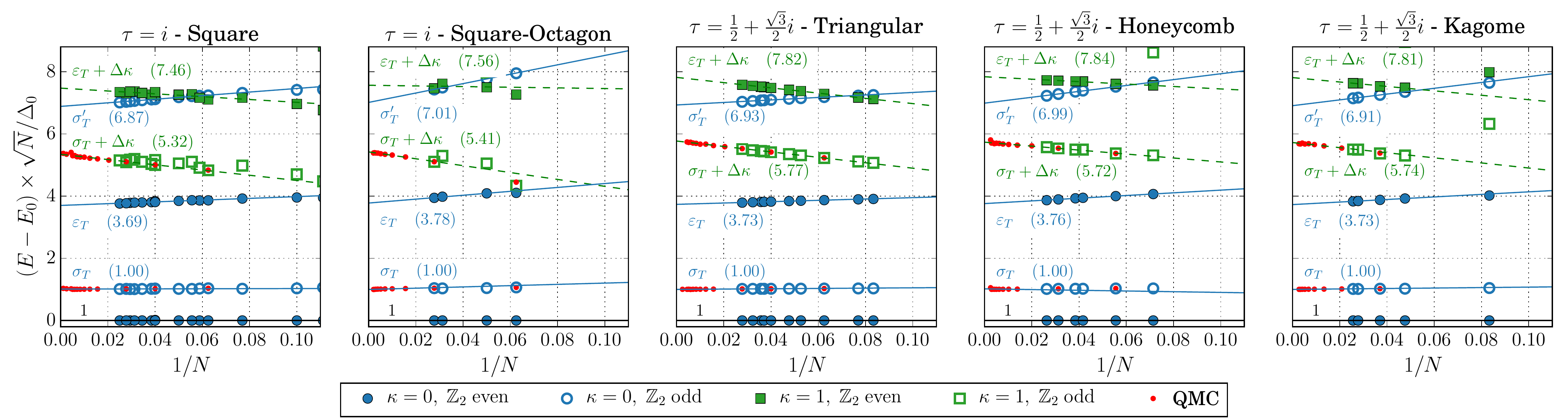}}
 \caption{\label{fig:TFI_Spectra}Normalized low-energy torus spectrum for the Ising QFT for the modular parameters $\tau=i$ and $\tau=1/2+\sqrt{3}/2i$ obtained with ED (large symbols) and QMC (small red filled circles). Filled (empty) symbols denote $\mathbb{Z}_2$ even (odd) levels. Linear fits in $1/N$ for levels with $\kappa=0$ ($\kappa=1$) are shown by blue solid (green dashed) lines (cf.~color coding in \figref{fig:TFI_geometry}) and the values of the fields after extrapolation to the thermodynamic limit $1/N \rightarrow 0$ are given in parentheses. The normalization constant $\Delta_0$ is chosen such that the first $\mathbb{Z}_2$ odd level extrapolates to one. We observe a universal torus spectrum for the lattices with the same type of IR cutoff (same $\tau$).} 
 
\end{figure*}
%%%%%%%%%%%%%%%%%%%%%%%%%%%%

%%%%%%%%%%%%%%%%%%%%%%%%%% TFI MODEL %%%%%%%%%%%%%%%%%%%%%%%%%%%%%%%%
\paragraph{3D Ising universality class ---} %: Transverse Field Ising Model ---} 
In order to demonstrate the universal nature of the low-energy spectrum we study the 2+1D transverse field Ising (TFI) model
\begin{equation}
H_\mathrm{TFI}=-J\sum_{\langle i,j\rangle} \sigma^z_i \sigma^z_j - h \sum_i \sigma^x_i 
\label{eqn:TFI}
\end{equation}
on five different two-dimensional Archimedian lattices \footnote{See Supplemental Material for a definition of the Archimedian lattices.}  
at their respective quantum critical point~\cite{Bloete2002}\footnote{We have computed the critical point for the Square-Octagon lattice as $(h/J)_c = 2.087(7)$ using a continous-time QMC algorithm similar to that of~\cite{Bloete2002}.}. 
In our finite size simulations the spatial setup is a torus whose linear extents are determined by two spanning vectors 
$\omega_1$ and $\omega_2$ (c.f. left part of Fig.~\ref{fig:TFI_geometry}). 
The finite area leads to a discrete momentum space (c.f. right part of Fig.~\ref{fig:TFI_geometry}) and is equivalent to an infrared cutoff in the field theory. 
The use of a lattice model on the other hand leads to an ultraviolet (UV) cutoff in the form of a Brillouin zone. 
In the following we will only consider tori with $L=|\omega_1|=|\omega_2|$ and two different choices of the modular parameter $\tau=\omega_2/\omega_1$:
$\tau=i$ $(\tau=1/2+\sqrt{3}/2i)$ corresponding to a square (hexagonal) symmetry. The square and square-octagon (triangular, honeycomb and
kagome) lattices are simulated using a square (hexagonal) IR-cutoff geometry to preserve the microscopic $C_4$ ($C_6$) point group symmetry
in the IR.

In a first step we have calculated the low-energy spectrum of the Hamiltonian~\eqref{eqn:TFI}
using exact diagonalization (ED) in all symmetry sectors on finite samples with up to $N=40$ spins in total. 
The spectrum can be divided into $\mathbb{Z}_2$ even and odd sectors (spin-flip symmetry), combined with irreducible representations of the lattice space group.
In the paramagnetic phase at large $h/J$ one finds a unique $\mathbb{Z}_2$ even ground state in the fully symmetric
spatial representation, with a finite gap above the ground state. At small $h/J$ one finds two quasi-degenerate ground states
in the $\mathbb{Z}_2$ even and odd sector respectively (both in the symmetric spatial representation), again with a finite gap 
above the ground state. At the quantum critical point $(h/J)_c$ however the low-lying spectrum collapses as $1/\sqrt{N} \sim 1/L$,
i.e. it exhibits a mass spectrum with the mass scale set by the IR cutoff. 
%\comment{
To eliminate this scaling we will multiply the excitation gaps with $\sqrt{N}$
in the following and will call that the spectrum.
%} 
In Fig.~\ref{fig:TFI_Spectra} we display the finite size
spectra at the Ising critical point for all five different lattices in the zero momentum sector $\Gamma=(0,0)$, as well as the first 
momentum away from the $\Gamma$ point ($\kappa=1$ in the right part of Fig.~\ref{fig:TFI_geometry}). 
Since the speed of light is not known at this stage, the spectrum for each lattice has been globally rescaled 
such that the extrapolated energy of the first excited level (which is $\mathbb{Z}_2$ odd and spatially symmetric) is set to one. One explicitly 
observes that the critical energy spectra of lattices with the same type of IR
cutoff $\tau$ (the two leftmost panels and the three rightmost panels) agree to
rather high precision with each other, when taking $1/N$ finite-size
corrections into account~\footnote{See Supplemental Material for a motivation of this $1/N$ finite-size extrapolation approach.}. 
This means that - as is generally expected from a field theory point of view - the obtained critical energy spectra indeed do not depend on the chosen UV 
discretization. In order to corroborate the extrapolations based on ED we performed 
extensive Quantum Monte Carlo (QMC) simulations~\cite{Bloete2002} of the transverse field Ising model at the critical point for all five lattices. 
Based on imaginary time spin-spin correlations it is possible to access the finite size gaps on lattices up to $N=30\times30$ lattice sites
\footnote{See Supplemental Material for further details about the used gap estimation procedure for QMC.}. 
These data points (red small filled circles) in Fig.~\ref{fig:TFI_Spectra} reproduce the ED data where available, and allow us to confirm and sharpen the precision of the  
extrapolated energy spectrum. Based on the quantum numbers of the first few low-lying energy levels we choose to label them as torus analogues of the spectrum of scaling 
dimensions of the 3D Ising CFT: $\sigma_T$ and $\sigma'_T$ refer to the first two levels in the $\mathbb{Z}_2$ odd sector in the spatially
symmetric representation, while $\epsilon_T$ is the first excited state (above the vacuum $1$) in the $\mathbb{Z}_2$ even and spatially symmetric
sector. The "$\ldots+\Delta\kappa$" label refers to levels at the first momentum away from the $\Gamma$ point, $\kappa=1$.
These levels are four-fold degenerate on the square torus, while they are six-fold degenerate for the hexagonal torus. 
Although there is no known relation between the torus spectrum and the scaling dimensions in flat space,
this phenomenological approach shows a qualitatively similar structure as the operator content of the quantum field theory.
%%%%%%%%%%%%%%%%%%%%%%%%EPSILON EXPANSION%%%%%%%%%%%%%%%%%%%%%%%%%%%%%%%%%%
\paragraph{$\epsilon$-expansion ---} 
We also compute the energy levels using $\epsilon$-expansion. Our starting point is $\phi^4$ theory, which we define by the Hamiltonian density
\beq
\mathcal{H} = \int d^dx \left[\frac{1}{2}\Pi^2 + \frac{1}{2}( \nabla \phi)^2 + \frac{s}{2} \phi^2 + \frac{u}{4!} \phi^4 \right] \label{onh}
\eeq
in $d$ dimensions with the equal-time commutator $[\phi(x,t),\Pi(x',t)] = i \delta^d(x - x')$, and specialize to the critical point, $s = s_c$, $u = u^\ast$. We generalize the two-dimensional torus to arbitrary dimension by taking $d/2$ copies of the desired tori in Fig.~\ref{fig:TFI_geometry}, so that all spatial point-symmetries are  preserved during the calculation and no extra length scales are introduced.

Our approach to the critical theory in a finite volume originated from L{\"u}scher \cite{L82}, and was extended to deal with finite size criticality in classical systems by others \cite{EBJZ85,RGJ85}. The key observation is that the zero mode of the field generates incurable infrared divergences in perturbation theory, so it must be separated and treated non-perturbatively. In the context of the finite-size spectrum, this can be understood from \eqref{onh} by noticing that the Gaussian theory at $s = 0$ does not contain any potential term for the zero mode, giving a continuous spectrum, whereas any finite $u$ will confine the zero mode producing a discrete spectrum. Therefore, the correct perturbative approach is to treat the momentum of the zero mode at the same order as its interactions.

% We split the fields into
% \bea
% \phi(x) &=& \mathcal{A}^{\frac{1-d}{4}}\varphi + \chi(x) \nn
% \Pi(x) &=& \mathcal{A}^{-\frac{d+1}{4}}  \pi + p(x)
% \eea
% where the zero mode terms have been normalized such that they are dimensionless and satisfy the commutation relation $[\varphi,\pi] = i$. The fields $\chi(x)$ and $p(x)$ only have finite-momentum modes in their Fourier series, and can be expanded in creation and annihilation operators. The Hamiltonian can then be written as
% \bea
% \mathcal{H} &=& \mathcal{H}_0 + V \nn
% \mathcal{H}_0 &=& \sum_{k \neq 0} |k| \left( b^{\dagger}(k)b(k) + \frac{1}{2} \right), \label{hsplit}
% \eea
% where $\mathcal{H}_0$ describes the Fock spectrum of the finite momentum modes, and we will subtract the total ground state energy. The momentum sum is over the reciprocal lattice vectors of the torus, and the interaction Hamiltonian $V$ contains all terms involving the zero mode and nonlinearities.
% %, which are given explicitly in Appendix \ref{}.
By splitting the fields in \eqref{onh} and proper normalization of the zero-mode terms the Hamiltonian can be decomposed into a quadratic part $\mathcal{H}_0$ describing the Fock spectrum of the finite-momentum modes, and an interaction part $V$ containing all zero-mode contributions and non-linearities.

At zeroth order, our states are given by finite momentum Fock states multiplied by arbitrary functionals of the zero mode, so these states are infinitely degenerate.
We then derive an effective Hamiltonian within each degenerate subspace using a perturbation method due to C. Bloch \cite{B58}. 
This effective Hamiltonian acts in a degenerate subspace, but its eigenvalues correspond to the exact eigenvalues of the original Hamiltonian to desired order.
It turns out, that the effective Hamiltonians take the form of a strongly-coupled oscillator with coefficients depending on the degenerate subspaces. The coefficients of the more complicated expansion for the energy levels (expansion in $\epsilon^{1/3}$) can be found in \cite{SCZ99}. In addition, the effective Hamiltonian will couple different Fock states with the same energy and momentum whenever possible, leading to off-diagonal terms. 
These off-diagonal terms were computed numerically from the unperturbed wave-function. 
Further details about the $\epsilon$-expansion approach can be found in the Supplemental Material~\footnote{See Supplemental Material, which includes Refs.~\cite{B58,K74,ZJ02,EBJZ85,WS16}}.

\begin{figure}[b]
 \centering
 \includegraphics[height=5cm]{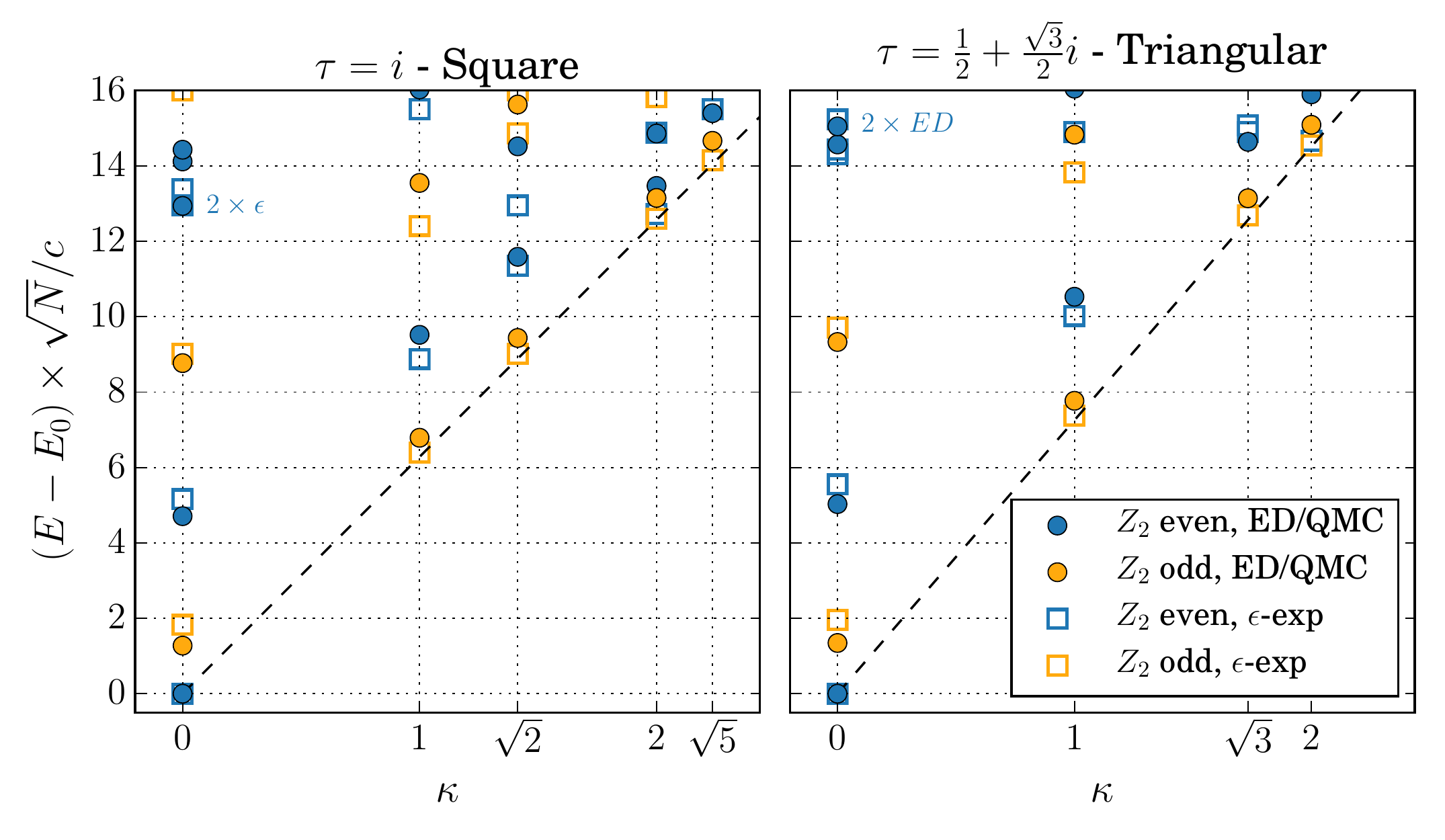}
 \caption{\label{fig:universal_spectrum} Universal torus spectra for the Ising QFT for the modular
 parameters $\tau=i$ (left panel) and $\tau=1/2+\sqrt{3}/2i$ (right panel). Full symbols
 denote numerical results obtained by ED and QMC (the lowest $Z_2$ odd levels), while empty symbols denote the
 $\epsilon$-expansion results. The dashed line shows a dispersion according to the speed of light.
 }
\end{figure}

In \figref{fig:universal_spectrum} we show the universal torus spectrum obtained from $\epsilon$-expansion for the two choices of $\tau$ and compare it to numerical results from ED and QMC computations~\footnote{See Supplemental Material for a listing of the complete low-energy torus spectra for the Ising transition from numerics and $\epsilon$-expansion.} normalized by the speed of light $c$~\cite{Sen2015}\footnote{See Supplemental Material, which includes Refs.~\cite{Sen2015,Hamer2000} for the details on the determination of $c$.}. We observe a remarkable agreement between the two different methods.
This further illustrates the interpretation of the torus spectra as a universal fingerprint of the critical field theory and their accessability from numerical finite lattice simulations.
%\comment{Footnote to add: We omit the use of error bars here. An accurate estimation of the errors is a difficult task since the fitting error alone is not taking further order corrections of the extrapolation into account. Nevertheless, for the lowest states for each $\kappa$ we expect the errors of the numerical results to be smaller than the plotted symbols in \figref{fig:universal_spectrum}. Another way to estimate the accuracy of the numerical results is to compare the values for the fields within the lattices of same IR cutoff $\tau$. However, one should consider, that the extrapolations for the square and triangular lattices are more precise than for the other lattices as the largest number of system sizes could be considered for those.}
The larger discrepancies between numerical and $\epsilon$-expansion data for some higher levels in the spectrum may result from the extrapolation to the thermodynamic limit using only ED data with strong finite-size effects, especially for $\kappa>0$~\footnote{For further studies it is worth noticing that $\epsilon$-expansion tends to overestimate the $\kappa=0$ levels while levels with $\kappa>0$ are commonly underestimated.}.

%%%%%%%%%%%%%%%%%%%%%%%%%%%%
%%%%%%%%%%%%%%%%%%%%%%%%%%%%

%%%%%%%%%%%%%%%%%%%%%%%%%%%%%
%\begin{figure}[h!]
% \centering
% \includegraphics[width=0.8\linewidth]{TFI_diff_gaps_speedoflight_sigma}
% \caption{\label{fig:sigma_shift} Shift of the $\sigma_T$ field in natural units.}
%\end{figure}
%%%%%%%%%%%%%%%%%%%%%%%%%%%%%

%%%%%%%%%%%%%%%%%%%%%%%%%% TORIC CODE %%%%%%%%%%%%%%%%%%%%%%%%%%%%%%%%
\paragraph{2+1D Ising* universality class ---} 
In this section we are investigating the confinement transition of a $\mathbb{Z}_2$ spin liquid. Such a topological quantum phase transition is characterized by the lack of any local order parameters. 
$\mathbb{Z}_2$ spin liquids are characterized by the presence of two bosons, the $e$ and $m$ particles. These fractionalized particles can only be created in pairs and obey mutual anyonic statistics. The confinement transition can then be driven by condensing either the $e$ or the $m$ particles. Without loss of generality, we will consider the condensation of the $m$ particles and call it's corresponding field $\phi$.
%\comment{$\mathbb{Z}_2$ spin liquids are characterized by their two types of {\em fractionalized} excitations, the $e$ and $m$ particles, that are bosons but obey mutual anyonic statistics. Each of those are only created in pairs and are deconfined. The confinement transition can equivalently be driven by condensing either of those particles, and without loss of generality, we will consider the condensation of the $m$ particles. In the following we call their corresponding field $\phi$.}
The critical theory turns out to be that of {\em Ising*}: $\phi$ can only be created in pairs, so the effective Lagrangian must be even in a real field $\phi$, implying we should only include $\mathbb{Z}_2$ even states in a critical Ising theory. In addition, $\phi$ and $-\phi$ are physically indistinguishable, and so both periodic and anti-periodic boundary conditions have to be considered.
We emphasize that this mapping is independent of any specific microscopic lattice model and should hold generically between universal theories and their topological counterparts.

As a microscopic model illustrating this transition we study the critical energy spectrum
of the Toric Code Hamiltonian perturbed by a longitudinal field~\cite{Trebst2007,Vidal2009,Tupitsyn2010,Dusuel2011,Wu2012}:
\begin{eqnarray}
\label{eq:ToricCode}
H_{TC} &= -J \sum_s A_s - J \sum_p B_p -h \sum_i \sigma_i^x\\
&A_s = \prod_{i \in s} \sigma_i^x, \; B_p = \prod_{i \in p} \sigma_i^z \nonumber
\end{eqnarray}
The $\sigma_i$ describe $S=1/2$-spins on the $2N$ edges of a square lattice, $p$ denotes a plaquette and $s$ a star on the lattice.
All $A_s$ and $B_p$ commute with each other and so the model can be solved analytically for $h=0$ by setting all $A_s=1$ and all $B_p=1$~\cite{Kitaev2003}. On a torus the ground state manifold is, however, four-fold degenerate and can be characterized by the eigenvalues $\pm 1$ of Wilson loops winding around the torus. An $e$ ($m$) particle is described by setting $A_s=-1$ ($B_p=-1$) on a star (plaquette). The longitudinal field introduces a dispersion for the $m$ particles which finally condense and drive the phase transition at $h=h_c$ by confinement of the $e$ particles~\cite{Jalabert1991,SSMV99,Senthil2000,Trebst2007}.

The above considerations regarding the relationship between Ising and Ising* quantum field theories (QFT) can be made very explicit for the Toric Code.
The Toric Code \eqref{eq:ToricCode} in the sector without $e$ particles ($A_s=1$) can be exactly mapped to an even TFI model on the dual square lattice with $N$ sites, where only the even spin-flip sector is present~\cite{Trebst2007, Hamma2008, Carr2010}. The groundstate manifold, described by the eigenvalues of the Wilson loops, maps to both, periodic and anti-periodic boundary conditions of the Ising model \footnote{See Supplemental Material for a detailed discussion of the mapping}. In the following we will make use of this mapping to compute the finite-size torus spectrum of the {\em Ising*} transition for $\tau=i$ using ED. 
%The relationship between Ising and Ising* theories is, however, very general and does not rely on this specific microscopic mapping.
%(the above sentence was repititive)

\begin{figure}
% % \subfigure[\label{fig:IsingStarA}]{
%   \includegraphics[height=5cm]{IsingStar_low_fields_ED}
% % }
% % ~
% % \subfigure[\label{fig:IsingStarB}]{
%   \includegraphics[height=5cm]{TC_Numerics_Epsilon_IR}
% % }
 \includegraphics[height=5cm]{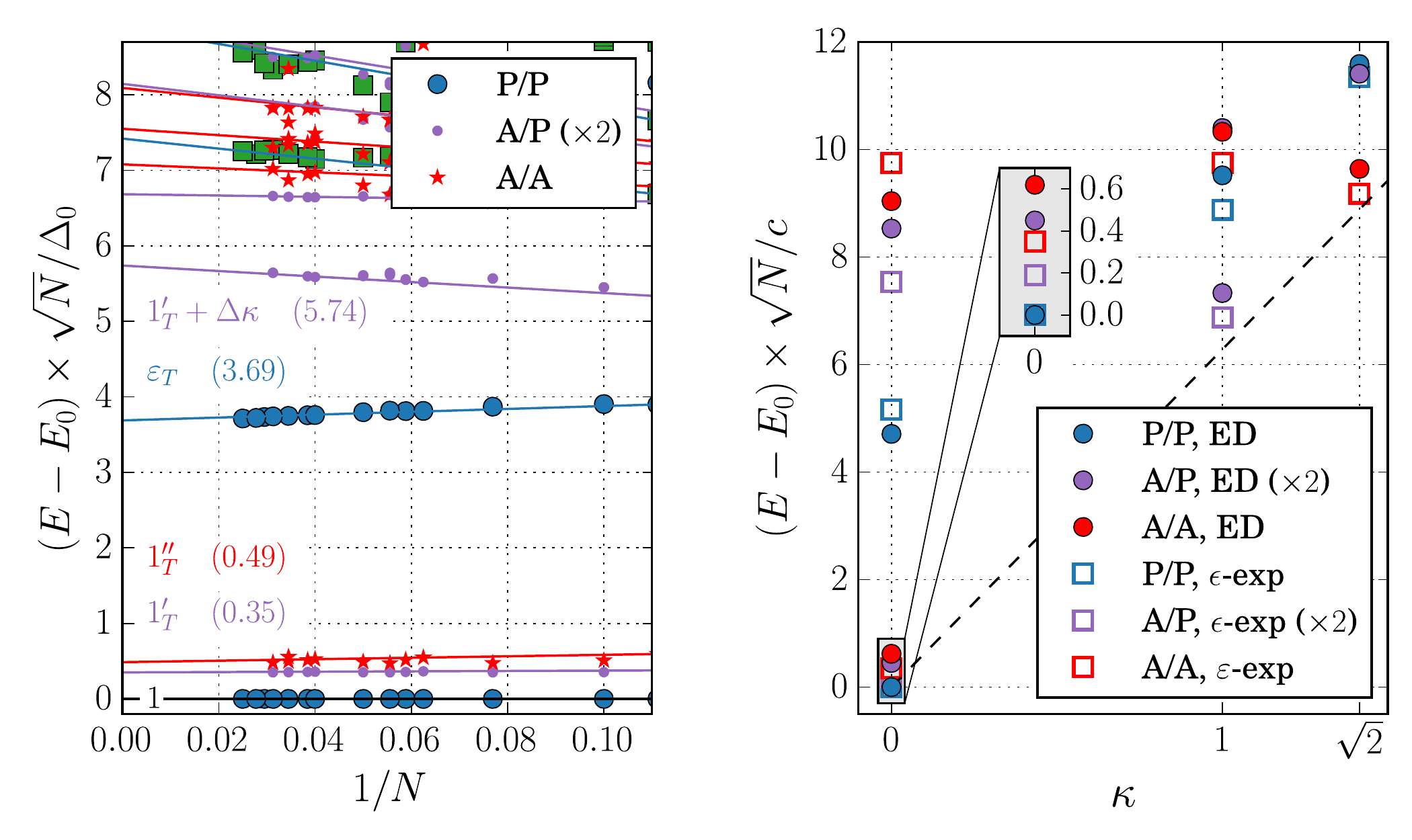}
 \caption{Universal torus spectra for the Ising* QFT and the modular
 parameter $\tau=i$. The labels A/P etc. denote the boundary conditions along the two directions of the torus, where P(A) means \mbox{(anti-)periodic}. Left: Normalized low-energy spectrum from ED with the same normalization constant $\Delta_0$ as in \figref{fig:TFI_Spectra}. The levels in the P/P sector are the $\varepsilon_T(+\Delta \kappa)$ levels from the TFI spectrum. A very remarkable feature is the presence of the four very low-lying levels which govern the four-fold degenerate groundstate manifold in the deconfined phase. See \figref{fig:TFI_Spectra} for further details. Right: Full symbols denote numerical results obtained by ED, while empty symbols denote $\epsilon$-expansion results. The dashed line shows a dispersion with the speed of light. The inset is a zoom into the four lowest levels. See \figref{fig:universal_spectrum} for further details.}
 \label{fig:IsingStar}
\end{figure}

In the left part of \figref{fig:IsingStar} we present the low-energy finite-size spectrum of the {\em Ising*} transition obtained with ED simulations. The spectrum is rescaled with the same factor $\Delta_0$ as in \figref{fig:TFI_Spectra} such that they can be easily compared. The relationship between the critical {\em Ising} and {\em Ising*} theories results in the fact that the levels called $\varepsilon_T (+\Delta \kappa)$ in \figref{fig:TFI_Spectra} are identically present in the {\em Ising*} spectrum (c.f.~P/P levels in \figref{fig:IsingStar}). The most remarkable feature, however, is the presence of very low-lying levels in the spectrum. They arise from the groundstate manifold in the spin-liquid phase, where their splitting exponentially scales to zero with $L$. At criticality they, however, scale as $1/\sqrt{N}$ as the entire low-energy spectrum. The small relative splitting of the four lowest levels is surprisingly small. 
%\comment{These levels are closely related to the anomalous critical exponent for the correlation functions $\eta$ at topological critical points due to fractional excitations \cite{Isakov2012a}. The unaltered levels in the spectrum are, on the other hand, related to the unchanged critical exponent $\nu$ which governs the divergence of the correlation length}. 
The right panel of \figref{fig:IsingStar} shows a comparison of the universal torus spectrum for an {\em Ising*} transition obtained with ED and $\epsilon$-expansion similar to \figref{fig:universal_spectrum}~\footnote{See Supplemental Material for a listing of the complete low-energy torus spectra for the Ising* transition from numerics and $\epsilon$-expansion.}. A zoom into the conspicuous low-energy levels is shown in the inset. 
Again we observe a decent agreement of the different methods. 
%This explicitely demonstrates once more that the effect of a neighbouring $\mathbb{Z}_2$ topological phase on the critical torus spectrum is a general one. It is not based on the exact mapping of the Toric Code considered here as a special case for an Ising* transition for numerical simulations.

%%%%%%%%%%%%%%%%%%%%%%%%%%%%
%%%%%%%%%%%%%%%%%%%%%%%%%%%%

%%%%%%%%%%%%%%%%%%%%%%%%%% BIBLIOGRAPHY %%%%%%%%%%%%%%%%%%%%%%%%%%%%%%%%
%\clearpage

\paragraph{Conclusions ---} 
We have computed the universal torus energy spectrum for the Ising and Ising* transitions in 2+1D 
providing a characteristic fingerprint of the corresponding conformal field theories 
and have highlighted the implications of a neighbouring $\mathbb{Z}_2$ spin liquid on the torus spectrum.
Additionally, we have highlighted a phenomenological picture based on the quantum numbers of the individual energy levels
which shows a structure qualitatively similar to the operator content of the field theory in flat space. 
Using the numerical and analytical technology presented in this paper it will be possible to inspect and chart the characteristic 
spectrum of more complex quantum critical points, such as $O(N)$ Wilson-Fisher fixed points, Gross-Neveu-Yukawa 
type phase transitions in interacting Dirac fermion models~\cite{Wang2014,Li2015} or designer Hamiltonians displaying deconfined 
criticality~\cite{Sandvik2007}. 

\acknowledgements
A.M.L. thanks R.C. Brower, J.L. Cardy and A.W. Sandvik for discussions.
L.-P.H. and M.S. acknowledge support through the Austrian Science Fund SFB FoQus (F-4018).
S.W. and S.S. are supported by the U.S. NSF under Grant DMR-1360789.
We thank A. Wietek for his help on computing large-scale ED results.
The computational results presented have been achieved in part using the Vienna Scientific Cluster (VSC).
This work was supported by the Austrian Ministry of Science BMWF as part of the UniInfrastrukturprogramm of the Focal Point Scientific Computing at the University of Innsbruck. 
Research at Perimeter Institute is supported by the Government of Canada through Industry Canada 
and by the Province of Ontario through the Ministry of Research and Innovation.   
This research was supported in part by the National Science Foundation under Grant No. NSF PHY11-25915.

%\bibliographystyle{apsrev4-1} //do not use a style for longbibliography
%\bibliographystyle{unsrt}
%\bibliography{TFI-QFT_abbrv,epsilon-expansion}
\input{IsingIsingStar.bbl}

\clearpage
\appendix
\begin{center}
\textbf{Supplemental Material: Universal Signatures of Quantum Critical Points from Finite-Size Torus Spectra}
\end{center}
\section{Lattice geometries}
\label{app:lattices}

\begin{figure}[h!]
 \centering
  \includegraphics[width=\columnwidth]{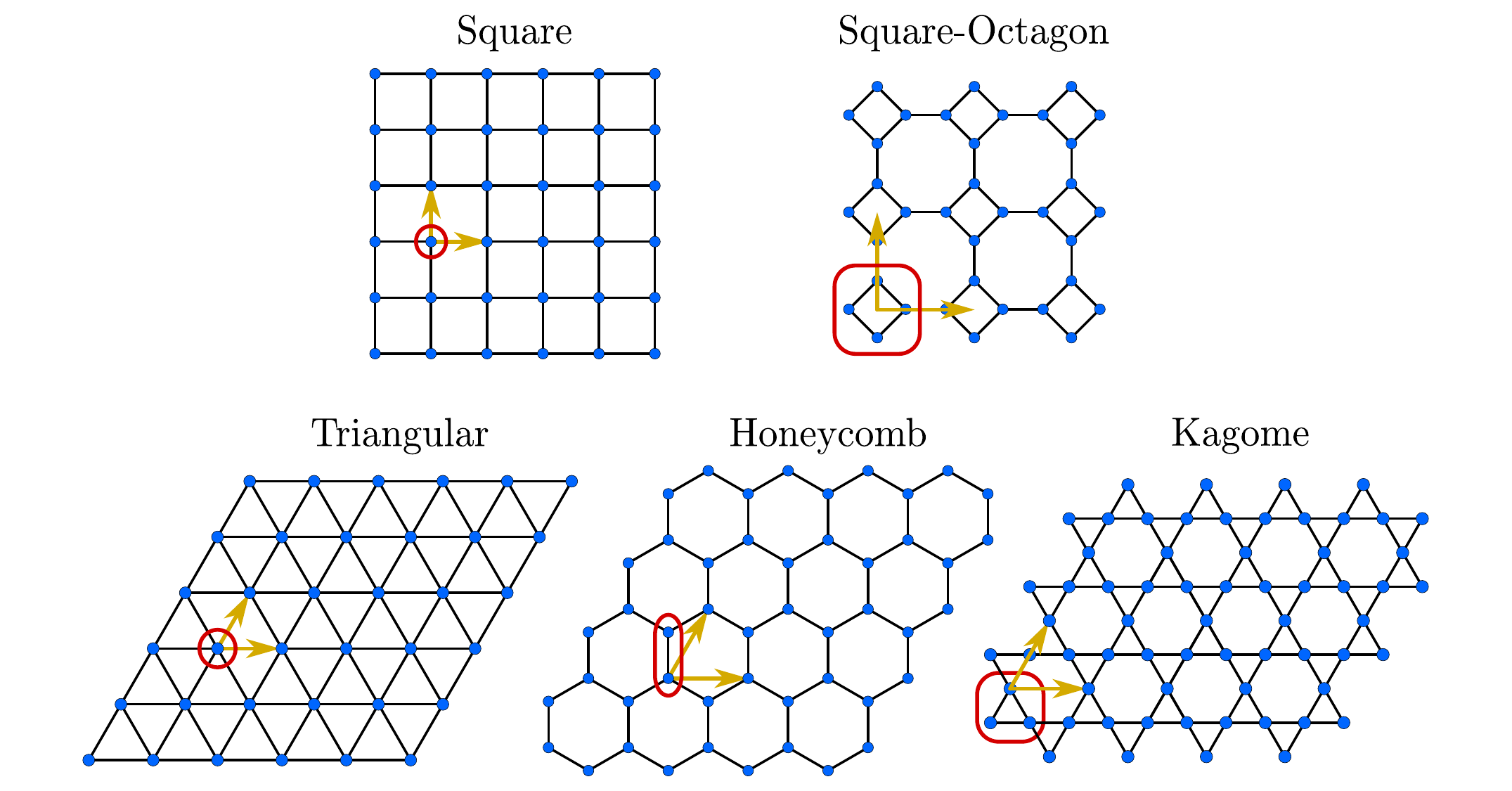}
 \caption{\label{fig:lattices}The different lattice geometries used for the TFI model. The red boxes indicate the lattice basis cells, the arrows mark the Bravais-vectors. The square and square-octagon lattices obey a $C_4$ rotational symmetry, the triangular, honeycomb and kagome lattices a $C_6$ rotational symmetry.}
\end{figure}

\section{Mapping the perturbed Toric Code onto the transverse field Ising model}
\label{app:tc_to_tfi}

\begin{figure}
 \centerline{\includegraphics[width=.6\columnwidth]{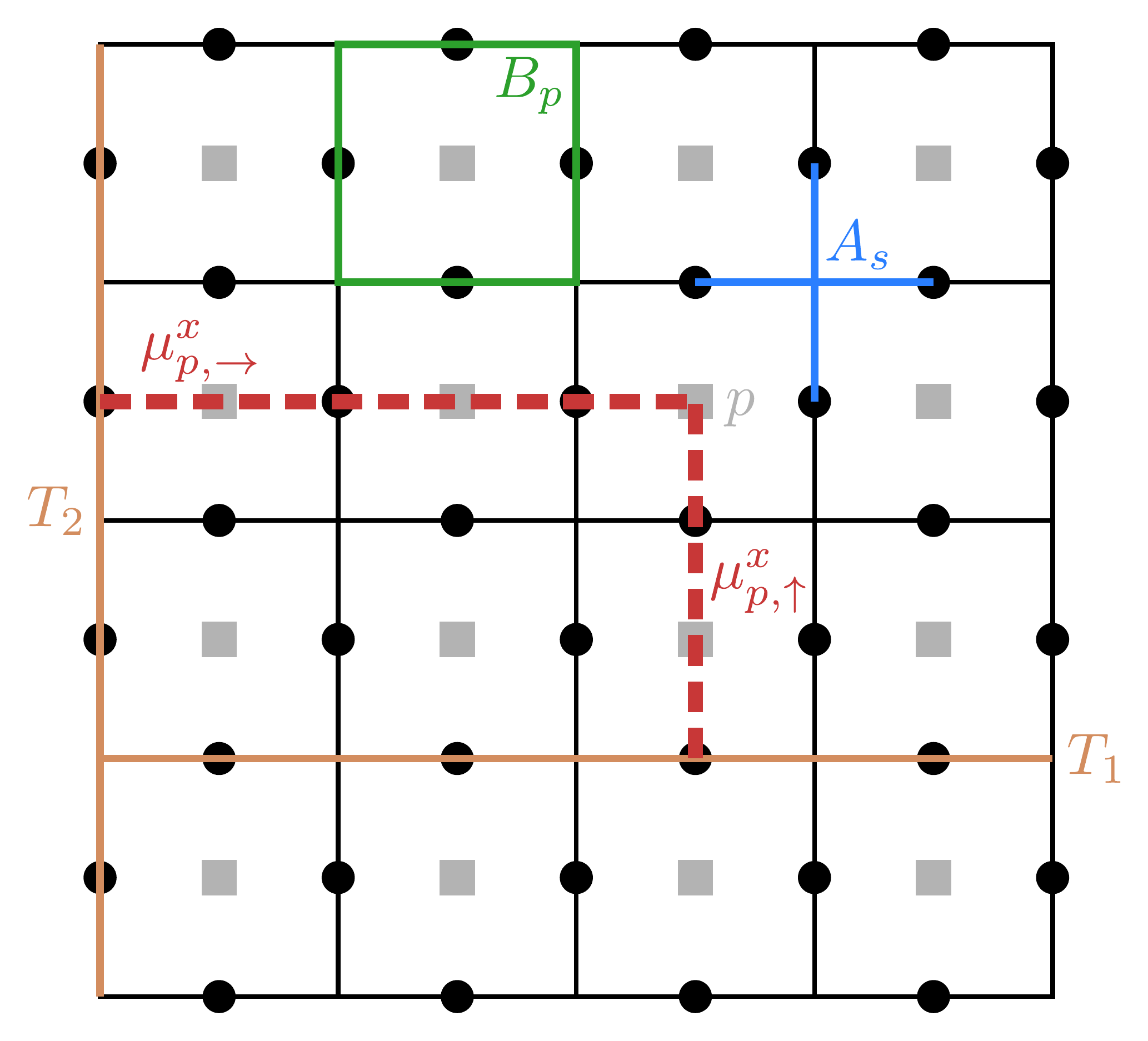}}
 \caption{\label{fig:TC_Sketch}The Toric Code on a torus. Black dots show the positions of the Toric Code variables $\sigma_i^{x,z}$, grey squares the dual lattice for the variables $\mu_p^{x,z}$. $T_{1,2}$ depict a choice of the two incontractible loops winding around the torus. See text for further details. }
\end{figure}

In this section, we demonstrate an exact mapping of the charge-free sector of the Toric Code model perturbed by a longitudinal field to a transverse field Ising model with only even states under spin-inversion. Such a mapping has already been used in previous studies of the Toric Code \cite{Trebst2007, Hamma2008, Carr2010}, here we will additionally show that the different groundstate sectors of the Toric Code  result in different boundary conditions of the transverse field Ising model.

The Hamiltonian of the Toric Code in a longitudinal field is given by 
\begin{align}
 H = -J &\sum_s A_s - J \sum_p B_p - h \sum_i \sigma_i^x\\
 A_s = &\prod_{i \in s} \sigma_i^x, \; B_p = \prod_{i \in p} \sigma_i^z
 \label{eq:TC}
\end{align}
where the $\sigma_i$ describe spins on the links of a square lattice, $p$ denotes a plaquette and $s$ a star on this lattice.
All $A_s$ and $B_p$ commute with each other and thus the GS of the Hamiltonian for $h=0$ can be found by setting  $A_s=1 \; \forall s$ and $B_p=1 \; \forall p$.
On a torus, however, not all of the $A_s$ and $B_p$ are linearly independent, as $\prod_s A_s=1$ and $\prod_p B_p=1$, leading to a 4-fould degenerate groundstate manifold.
This groundstate manifold can be distinguished by the expectation values of the Wilson loop operators $t_{1,2} = \prod_i \sigma_i^x$ where the paths wind around the torus along two non-contractible loops through the centers of the edges of the lattice (e.g. parallel to $T_{1,2}$ in \figref{fig:TC_Sketch}). 

To perform the mapping to a transverse field Ising model we first note, that $A_s$ and $t_{1,2}$ are still conserved for $h \neq 0$, when the longitudinal field is turned on. So, we consider the charge-free sector, $A_s=1 \; \forall s$, which describes the low-energy physics even at criticality, and define the new variables
\begin{align}
 \label{eq:Bpmap}
 \mu_p^z &= B_p \\
 \mu_{p,\rightarrow(\uparrow)}^x &= \prod_{i \in c_{p \rightarrow(\uparrow)}} \sigma_i^x
\end{align}
on each site $p$ of the dual lattice (center of plaquette $p$) \cite{Hamma2008}. We choose two incontractible paths $T_{1,2}$ in $\hat{x}(\hat{y})$ direction along the lattice. The path $c_{p \rightarrow(\uparrow)}$ is then a straight path from $T_{2(1)}$ to the site $p$ in $\hat{x}(\hat{y})$-direction along the dual lattice (cf.~\figref{fig:TC_Sketch}).
It is straightforward to show that these variables fulfill the Pauli-Algebra $\{\mu_p^x, \mu_p^z\}=0, (\mu_p^x)^2=1$ and that
\begin{align}
 \label{eq:tc_to_tfi1}
 \sigma_i^x(\hat{x}) &= \mu_{p(i),\uparrow}^x \mu_{p(i)-\hat{y},\uparrow}^x \\
 \label{eq:tc_to_tfi2}
 \sigma_i^x(\hat{y}) &= \mu_{p(i),\rightarrow}^x \mu_{p(i)-\hat{x},\rightarrow}^x
\end{align}
where $\sigma_i^x(\hat{x}(\hat{y}))$ describes a Pauli operator on a link in $\hat{x}(\hat{y})$-direction on the lattice.

With this, the TC eventually maps onto the well-known TFI model
\begin{equation}\label{eq:tfi_from_tc}
 H_{TFI} = -h \sum_{\langle p, q\rangle} \mu_p^x \mu_q^x - J_p \sum_p \mu_q^z + const.
\end{equation}
on the dual lattice and $A_s = 1 \; \forall s$, as it was imposed.

% So far we have not considered the mapping of the $A_s$ terms in the original model. Each $A_s$ term can be written as 
% \begin{equation}
%  A_s = \prod_{i \in s} \sigma_i^x = (\mu_{(j,k)}^x \mu_{(j,k+1)}^x) (\mu_{(j+1,k)}^x \mu_{(j+1,k+1)}^x) (\mu_{(j,k)}^x \mu_{(j+1,k)}^x) (\mu_{(j,k+1)}^x \mu_{(j+1,k+1)}^x) = 1.
% \end{equation}
% The last equality follows from the fact that each $\mu_a$ term comes as a square in the equation. Thus the presented duality of TC and TFI model holds only for the sector without any $A$-excitations in the TC.

The resulting transverse field Ising model \eqref{eq:tfi_from_tc} is invariant under global spin-inversion $\mathcal{I} = \prod_p \mu_p^z$. From \eqref{eq:Bpmap} it immediately follows that
\begin{equation}
 \mathcal{I} = \prod_p B_p = 1
\end{equation}
where the last equality is always satisfied on a torus and so the Toric Code maps to an {\em even} transverse field Ising model.

Let us finally apply the mapping on the different groundstate sectors characterized by the eigenvalues of $t_{1,2}$. Using \eqref{eq:tc_to_tfi1} and \eqref{eq:tc_to_tfi2} it follows that 
\begin{equation}
 t_1 = \prod_{p=0}^{L-1} \mu_{(p,j)}^x \mu_{(p+1,j)}^x = \mu_{(0,j)}^x \mu_{(L,j)}^x
\end{equation}
where the index $(p,j)$ labels the position $p \hat{x} + j \hat{y}$ on the dual lattice and $L$ is the linear extend of the torus. An equivalent relation can be computed for $t_2$. The different groundstate sectors of the Toric Code therefore map onto periodic and antiperiodic boundary conditions of the transverse field Ising model for both directions around the torus.
% $\mu_{(L,j)}^x = \pm \mu_{(0,j)}^x$

\section{Finite size gap estimation with QMC}
  To estimate the gaps in a larger range of system sizes than reachable by ED, we use a continuous-time world-line Monte-Carlo scheme, supplemented with a cluster update  \cite{Bloete2002} to overcome critical slowing down at the quantum phase transition.

  For our computations, we used system linear sizes ranging up to $L=30$ (for the simplest lattices). For each system, the average energy was computed from several (from 16 to 256) independent runs of $10^4$ to $10^6$ measurements. Between two measurements, the number of cluster updates $n_c$  was chosen so that $n_c\langle s_c\rangle\gtrsim \beta L^2$, where $\langle s_c\rangle$ is the average cluster size. This leads to an autocorrelation-time of order one Monte Carlo step or less.
  
    The world-line Monte-Carlo allows to extract the excitation spectrum of the system through the evaluation of the imaginary-time spin-spin correlation function. 
    Indeed, the spin-spin correlation function at momentum $\mathbf{q}$ is given by
    \begin{equation}
      \begin{array}{ll}
	S^{zz}(\mathbf{q},\tau)&=\dfrac{1}{\beta N}\left\langle\displaystyle\sum_{i,j}{\displaystyle\int_0^{\beta}{d\tau'\,e^{-\mathrm{i}{\bf q}\cdot({\bf r}_i-{\bf r}_j)}s_i^z(\tau')s_j^z(\tau'+\tau)}}\right\rangle\\
	&\underset{\beta\to\infty,\tau\to\infty}{\approx} e^{-\Delta_\mathbf{q} \tau},
      \end{array}
    \end{equation}
    where $\Delta_\mathbf{q}$ is the excitation gap at momentum $\mathbf{q}$.
    \begin{figure}[h!]
      \includegraphics[width=.45\textwidth]{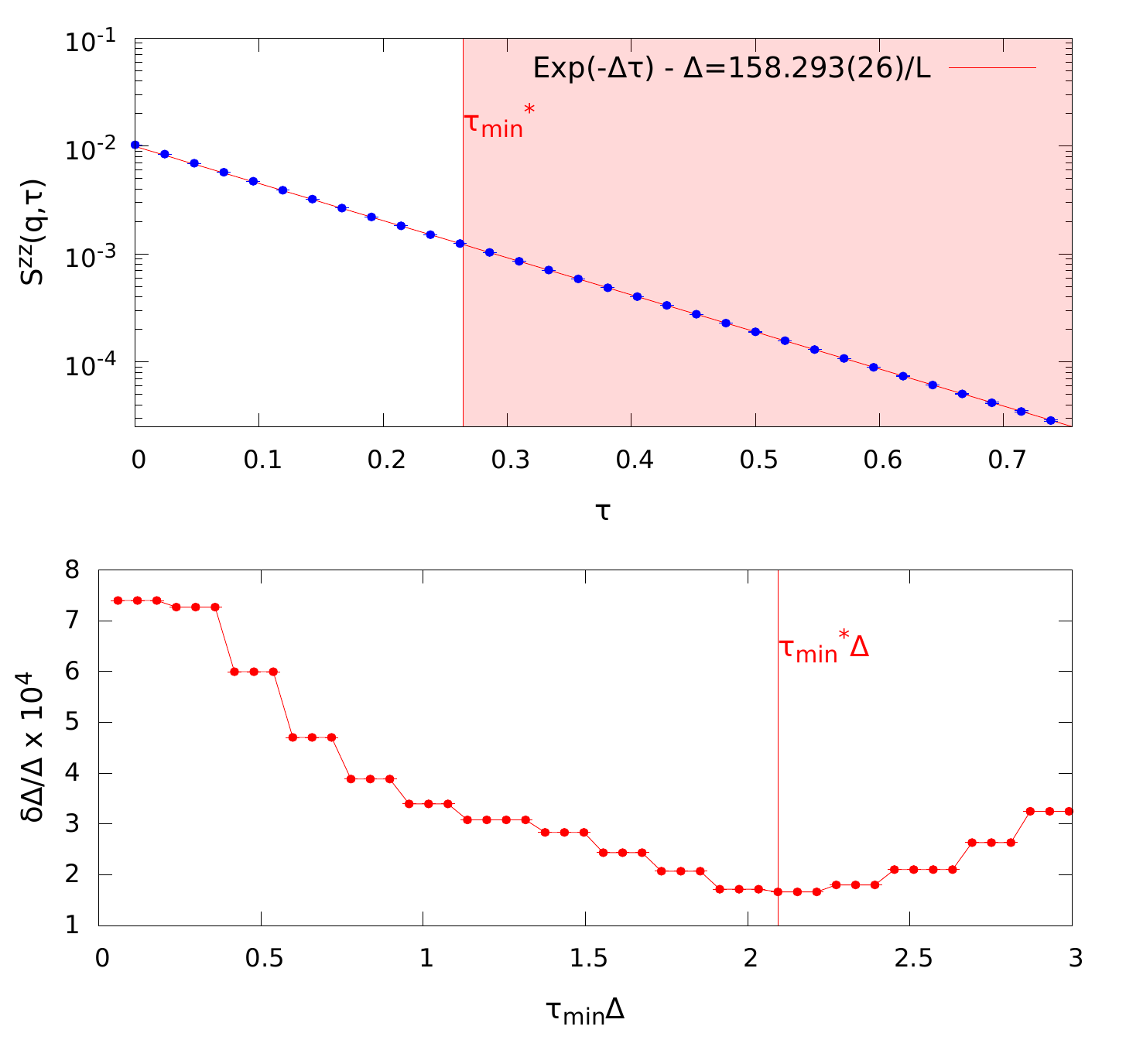}
      \caption{Extraction of the excitation gaps from QMC data (here we display the example of a triangular lattice system of size $L=20$, at the smallest non-zero momentum).
      {\sl top panel} Spin-spin correlation function as a function of imaginary time compared to the fitted exponential.
      {\sl bottom panel} Relative fitting error on the evaluation of the gap as a function of the minimal imaginary time used in the fit.
      }
      \label{GapEstimation}
    \end{figure}
    To optimize the estimation of the gap, we fit an exponential decay to the spin-spin correlation function for imaginary times $\tau>\tau_{\text{min}}$ (Fig.~\ref{GapEstimation}, top panel) 
    and find the value $\tau_{\text{min}}^*$ that minimizes the relative fitting error on the value of the gap (Fig.~\ref{GapEstimation}, bottom panel).

\section{Finite-size extrapolation of the energy gaps}

\begin{figure}[h]
 \centering
 \includegraphics[width=\columnwidth]{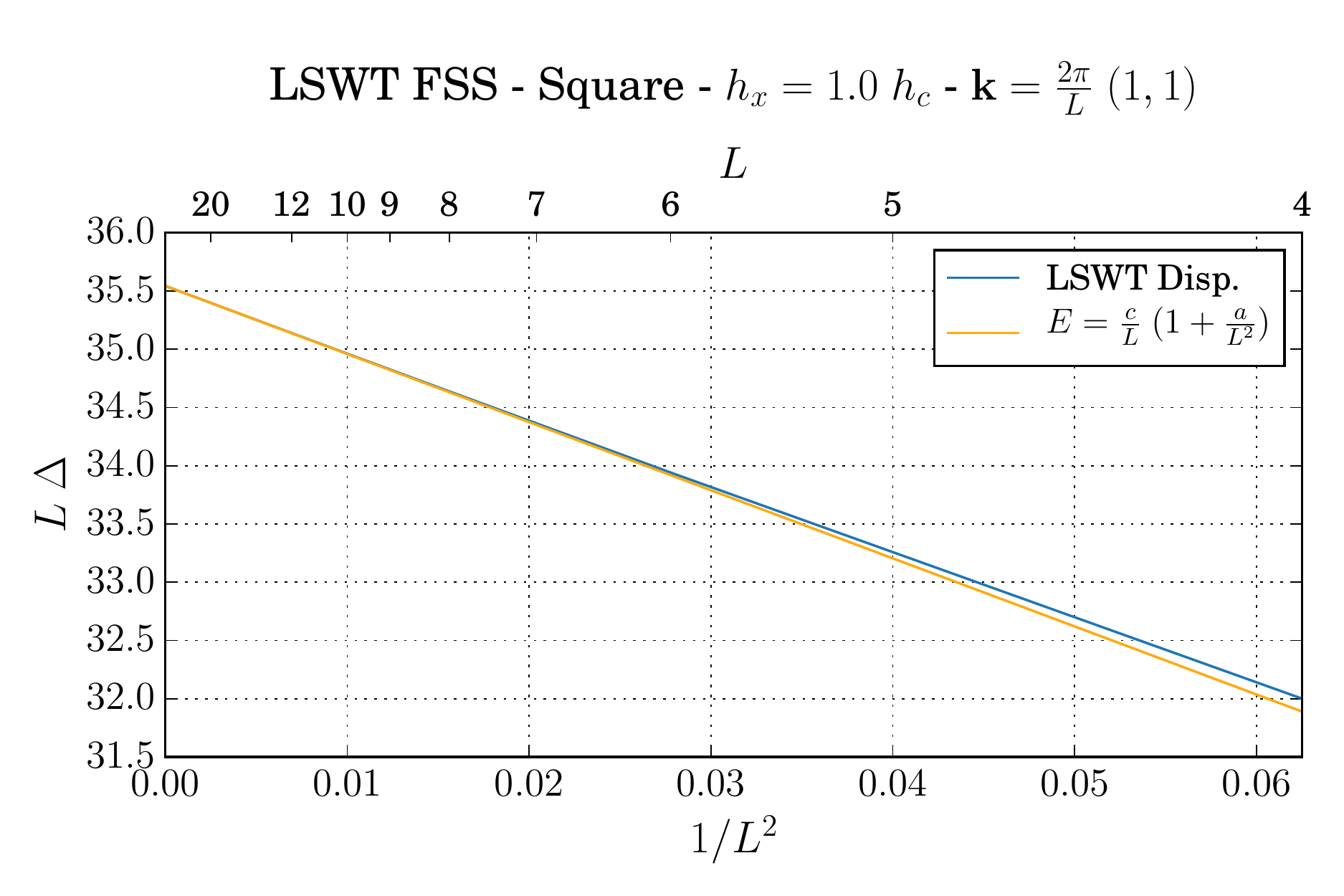}
 \caption{\label{fig:fsslswt} Finite size scaling of a $\kappa=\sqrt{2}$
     spectral level at criticality within a linear spin-wave theory (LSWT) approach. The
     blue curve is according to the LSWT dispersion, the yellow curve is a
     power-series of this dispersion up to the second non-vanishing order.}
\end{figure}

To motivate the dominant $1/N$ scaling used in our extrapolations of the
finite-size energy gaps $L \; \Delta$ we have calculated the dispersion relation
$\epsilon(\mathbf{k})$ at criticality within linear spin-wave theory (LSWT). We
can use this dispersion to compute the finite-size scaling of a $\kappa\neq0$
level within this approach by considering a momentum $\mathbf{k}=\frac{1}{L} (u,
v)$ (on the square lattice) and expanding the resulting expression in powers of
$1/L$
\begin{align}
    \epsilon(\mathbf{k}) &= \frac{c}{L} \left(1 - \frac{a}{L^2} +
\mathcal{O}(1/L^4)\right) \\
    c &= \sqrt{(u^2+v^2)/2} \\
    a &= \frac{u^4+v^4}{24(u^2+v^2)}
\end{align}
The dominant corrections to the spectral levels $L\;\Delta$ within the LSWT
approach are thus $1/L^2 = 1/N$. In \figref{fig:fsslswt} we compare the
finite-size extrapolation of a spectral level in the full LSWT approach and in
the power-series expansion up to $1/N$. The effect of higher order corrections
is small already for intermediate size systems.

\begin{figure}[h]
 \centering
 \includegraphics[width=\columnwidth]{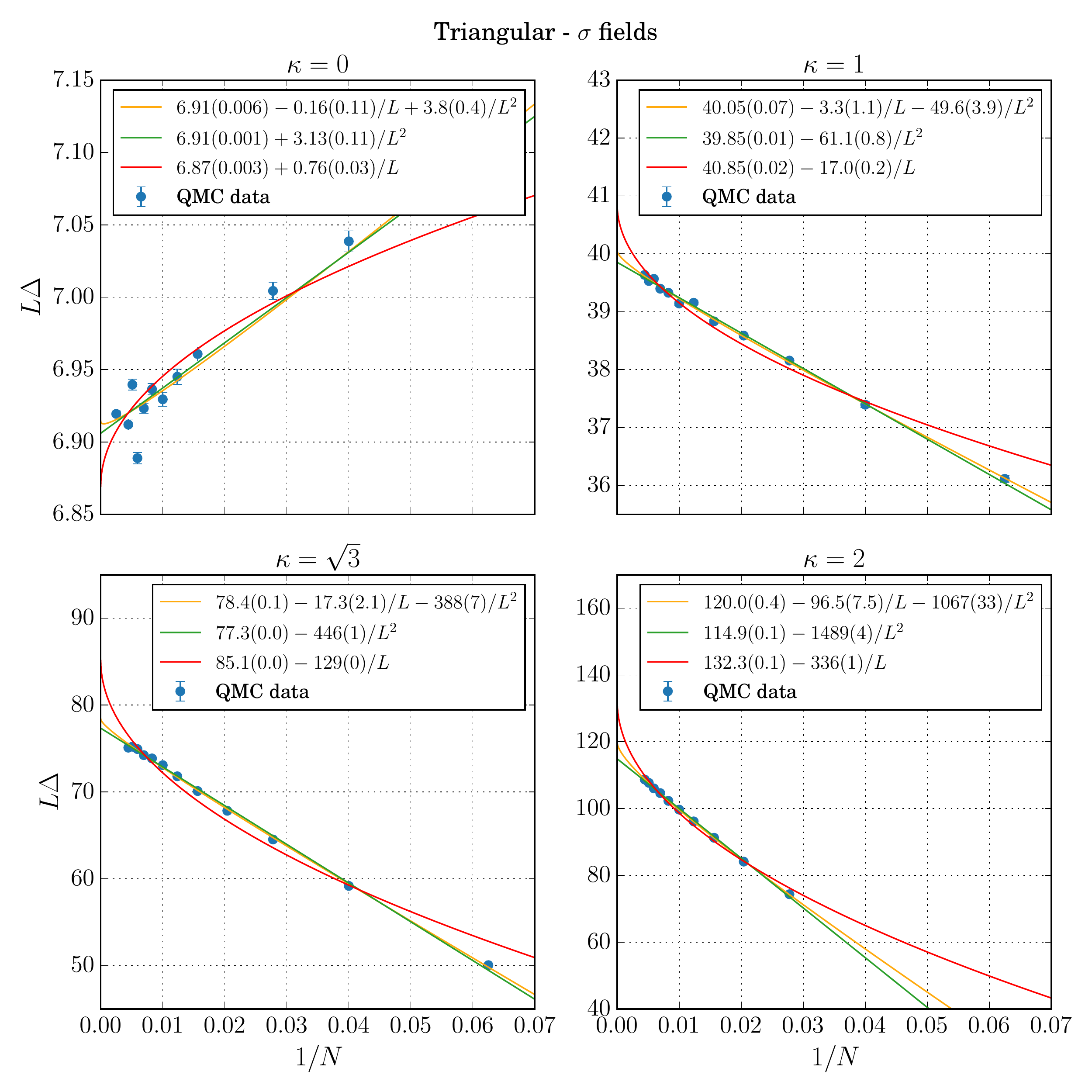}
 \caption{\label{fig:extrapolations}Comparison of different fitting approaches for the extrapolation of
     the finite-size $\sigma_T$ levels in different $\kappa$ sectors. The values
 in parantheses give the pure fitting error for the given values.}
\end{figure}

We have also considered fitting approaches with additional $1/L$ terms for the
extrapolation of the spectral levels to the thermodynamic limit $N\rightarrow
\infty$. In \figref{fig:extrapolations} we compare different fitting approaches
of the QMC gaps $L\;\Delta$ for $\sigma_T$ levels in different $\kappa$ sectors on the
triangular lattice. While a correction in purely $1/L$ (red fits) gives poor
results, correction terms in $1/L^2$ with and without an additional $1/L$ term
fit the data very well leading to similar extrapolated gaps. The fitting
procedure with both terms is, however, much more instable when very small
systems are included in the fit, often leading to strong dips close to $1/N=0$. 

\section{Speed of light from QMC}
  For each lattice, in order to extract the speed of light $c$, we proceed as follows. 
  We first extract with QMC the energies $E^{\sigma_T}_L(\kappa)$ of the lowest $\mathbb{Z}_2$ odd levels at a given system size $L$. 
  We then extrapolate those to the thermodynamic limit $E^{\sigma_T}(\kappa)$.  
%   We first extract the dispersion $\omega^{\sigma_T}_{\kappa}$ of the lowest $\mathbb{Z}_2$ odd levels by extrapolating to the thermodynamic limit the one obtained at finite size with QMC.
  We finally fit a line $E^{\sigma_T}(\kappa)=\delta E+c\cdot\kappa$ to this extrapolated data, in the interval $\left[\kappa_{\min},\kappa_{\max}\right]$~\cite{Sen2015}.
  Since we expect the levels at small momenta to be affected by the periodic boundary conditions, we take $\kappa_{\min}>0$. For large momenta, the finite-size curvature effects render the
  thermodynamic-limit extrapolation ill-defined, and therefore one has to introduce an ultraviolet cutoff $\kappa_{\max}$. There is ambiguity in the choice of $\kappa_{\min}$ and $\kappa_{\max}$,
  but for all {\it reasonable} choices 
  ({\sl i.e.} so that enough points lie within the linear regime in this interval), 
%   ({\sl i.e.} $1\leq \kappa_{\min} \leq 4$ and $\kappa_{\min} + 5\leq \kappa_{\max} \leq L_{\max}/4$, with $L_{\max}$ the largest attainable system size for this lattice), 
  one gets a value for $c$ with a fitting asymptotic standard error of less than 0.5\%. However, the value of $c$ thus obtained varies by about 1\% (2\%) for the square and triangular lattices
  (square-octagon, honeycomb and kagome lattices) across the various choices of fitting intervals. This leads to the speed of light estimates given in Tab.~\ref{tab:speedoflight}.

\begin{table}[h!]
 \centering
  \begin{tabular}{|c|c|c|}
  \hline
  {\bf Lattice} & $\mathbf{c/J}$ & (prev. works) \\\hline\hline
  Square & 3.323$\pm$0.033 & 3.01$\pm$0.09\cite{Hamer2000}\\\hline
  Square-Octagon & 5.126$\pm$0.103& \\\hline
  Triangular & 2.047$\pm$0.020& \\\hline
  Honeycomb & 2.923$\pm$0.058& \\\hline
  Kagome & 2.013 $\pm$ 0.040&\\\hline
  \end{tabular}
  \caption{\label{tab:speedoflight}Speed of light for each lattice geometry, from QMC.}
\end{table}

%%%%% EPSILON EXPANSION APPENDIX
\section{Spectrum of the Wilson-Fisher CFT on a torus: $\epsilon$-expansion}
\label{app:epsilon}

In this appendix we elaborate on the calculation of the spectrum from the $\epsilon$ expansion. A more detailed exposition which generalizes to the O(N) model and includes deviations from the critical point will be presented in a future publication. The Hamiltonian is
\beq
H = \int d^dx \left[\frac{1}{2}\Pi^2 + \frac{1}{2}( \nabla \phi)^2 + \frac{s}{2} \phi^2 + \frac{u}{4!} \phi^4 \right] \label{onhapp}
\eeq
where we will always tune to the critical point, $s = s_c$, $u = u^\ast$, in final expressions. The system is taken to be on $d/2$ copies of a 2-torus with modular parameter $\tau = \tau_1 + i \tau_2$ and area $\mathcal{A} = \mathrm{Im}\left( \omega_2 \omega_1^{\ast} \right) = \tau_2 L^2$. We will use complex coordinates, $x = x_1 + i x_2$, for each copy of the torus (see Fig.~1 in the main text).

As discussed in the main text, a gapless field theory on a finite volume leads to incurable infrared divergences due to the zero-momentum component of the fields. The solution to this problem is to split the fields into a zero-mode part and a finite momentum part,
\bea
\phi(x) &=& \mathcal{A}^{\frac{1-d}{4}}\varphi + \psi(x) \nn
\Pi(x) &=& \mathcal{A}^{-\frac{d+1}{4}}  \pi + p(x)
\eea
where the zero mode terms have been normalized such that they are dimensionless and satisfy the commutation relation $[\varphi,\pi] = i$. The fields $\chi(x)$ and $p(x)$ only have finite-momentum modes in their Fourier series:
\bea
\psi(x) &=& \frac{1}{\mathcal{A}^{d/4}} \sum_{k \neq 0} \frac{e^{i k \cdot x}}{\sqrt{2 |k|}}\left( b(k) + b^{\dagger}(-k) \right) \nn
p(x) &=& -\frac{i}{\mathcal{A}^{d/4}} \sum_{k \neq 0} \sqrt{\frac{|k|}{2}}e^{i k \cdot x}\left( b(k) - b^{\dagger}(-k) \right)
\eea
The momentum sums are over $d/2$ copies of the complex dual lattice to the torus, and the dot product is given by $k \cdot x = \mathrm{Re}\left( k x^{\ast} \right)$. With this decomposition, the Hamiltonian can be split up as
\beq
H = H_0 + V 
\eeq
with
\bea
H_0 &=& \sum_{k \neq 0} |k| b^{\dagger}(k)b(k) \nn
V &=&  \frac{1}{\sqrt{\mathcal{A}}}\left[ \frac{1}{2}\pi^2 + \frac{1}{2} \mathcal{A} s \varphi^2 + \frac{u \mathcal{A}^{\epsilon/2}}{4!} \varphi^4 \right] \nn
&+& \frac{u \mathcal{A}^{\epsilon/2}}{\sqrt{\mathcal{A}}} \frac{\varphi^2}{8} \sum_{\mathbf{k} \neq 0} \frac{\chi(-k)\chi(k)}{\mathcal{A}^{1/2} |k|}  \nonumber \\
&+& \frac{u \mathcal{A}^{\epsilon/2}}{\sqrt{\mathcal{A}}} \frac{\varphi}{6} \sum_{k,k' \neq 0} \frac{\chi(k) \chi(k') \chi(-k - k')}{(8 \mathcal{A}^{3/2} \omega_{k}\omega_{k'}\omega_{k + k'})^{1/2}}  \nn
&+& \frac{u \mathcal{A}^{\epsilon/2}}{\sqrt{\mathcal{A}}} \frac{1}{4!} \sum_{k_i \neq 0 \atop k_1 + k_2 = k_3 + k_4} \frac{\chi(k_1) \chi(k_2) \chi(k_3) \chi(k_4)}{4 (\mathcal{A}^2 |k_1||k_2||k_3||k_4|)^{1/2}} \qquad
\eea
where we define $\chi(k) \equiv b(k) + b^{\dagger}(-k)$. Here, $\mathcal{H}_0$ describes a Fock spectrum of finite momentum states, and the interaction Hamiltonian $V$ contains all terms involving the zero mode and nonlinearities. In this paper, we will always set the ground state energy to zero; a future publication will discuss the universal dependence of the ground state energy on $L$ and on relevant perturbations $s - s_c$.

At zeroth order, our states are given by finite momentum Fock states multiplied by arbitrary functionals of the zero mode, 
\beq
H_0 \Psi[\varphi]|k, k', \cdots \rangle = \left( |k| + |k'| + \cdots \right)\Psi[\varphi]|k, k', \cdots \rangle
\eeq
Since we can multiply by any normalizable functional $\Psi[\varphi]$, these states are infinitely degenerate. This degeneracy is broken in perturbation theory. We use a perturbation method due to C. Bloch which is well-suited to degenerate problems \cite{B58}. For a review of this method and its relation to other effective Hamiltonian methods, see Ref.~\cite{K74}. The main idea is to consider each degenerate subspace separately, but construct an effective Hamiltonian within each subspace whose eigenvalues give the exact energy. So if we consider a degenerate subspace of $H_0$,
\beq
H_0 |\alpha_0 \rangle = \epsilon_0 | \alpha_0 \rangle
\eeq
this perturbation method constructs a new operator $H_{eff}$ which acts on this subspace but gives the exact energy levels,
\beq
H_{eff}|\alpha_0 \rangle = E_{\alpha} |\alpha_0 \rangle \label{effdef}
\eeq
where $E_{\alpha} = \epsilon_0 + \mathcal{O}(V)$.

%The main idea is to consider a degenerate subspace of the unperturbed Hamiltonian,
%\beq
%H_0 | \alpha_0 \rangle = \epsilon_0 | \alpha_0 \rangle,
%\eeq
%where $\Omega_0 = \mathrm{Span}\left\{ | \alpha_0 \rangle \right\}$ denotes the space of all degenerate states. The exact Hamiltonian will split these states,
%\beq
%H| \alpha \rangle = E_{\alpha} | \alpha \rangle, \qquad E_{\alpha} = \epsilon_0 + \mathcal{O}\left( V \right).
%\eeq
%We can take the states $|\alpha \rangle$ to be orthonormal, and denote this space of states by $\Omega = \mathrm{Span}\left\{ | \alpha \rangle \right\}$. Then if the perturbation is small, there should exist a one-to-one correspondence between $\Omega_0$ and $\Omega$. If we let $P_0$ be the projection operator onto the space $\Omega_0$, then we can define the states
%\beq
%|\alpha_0 \rangle = P_0 | \alpha \rangle
%\eeq
%which span $\Omega_0$ (although they are not orthonormal). There also exists an operator $U$ such that
%\bea
%&&U | \alpha_0 \rangle = | \alpha \rangle, \nonumber \\
%&&U | \phi \rangle = 0 \quad \forall | \phi \rangle \notin \Omega_0.
%\eea
%Given the operator $U$, we can now define the effective Hamiltonian:
%\bea
%H_{eff} = P_0 H U \nn
%H_{eff} |\alpha_0 \rangle = E_{\alpha} |\alpha_0 \rangle. \label{effdef}
%\eea

The expression for $H_{eff}$ can be obtained perturbatively in $V$, which was the main result of Bloch's work \cite{B58}. To leading order, the effective Hamiltonian for a given degenerate subspace is given by
\beq
H_{eff} = \epsilon_0 P_0 + P_0 V P_0 + P_0 V \frac{1 - P_0}{\epsilon_0 - \mathcal{H}_0} V P_0 + \cdots
\label{eqn:effham}
\eeq
where $P_0$ is the projection operator onto the degenerate subspace of interest. At this order in perturbation theory the effective Hamiltonian is hermitian, although to higher orders one needs to make a unitary transformation insure hermiticity \cite{B58,K74}. 

As a definite example, we give the effective Hamiltonian acting on the Fock vacuum, $\Psi[\varphi] |0 \rangle$. From Eq.~(\ref{eqn:effham}), the effective Hamiltonian takes the form
\beq
H_{eff,k=0} = h_{k=0} | 0 \rangle \langle 0 | \nonumber
\eeq
\beq
h_{k = 0} = \langle 0 | V | 0 \rangle - \langle 0 | V \left( \frac{1 - | 0 \rangle \langle 0 |}{H_0} \right) V | 0 \rangle + \cdots
\label{eqn:hk0}
\eeq
When this acts on the ground state manifold, it generates a Schr\"odinger equation for the zero-mode functional,
\beq
h_{k = 0} \Psi[\varphi] = E \Psi[\varphi]
\eeq
We now obtain $h_{k=0}$ from evaluating the expectation values in Eq.~(\ref{eqn:hk0}). This involves UV divergent sums which requires renormalization, but the renormalization constants and RG equations will be identical to the infinite volume case as a consequence of finite-size scaling. We renormalize the theory using dimensional regularization with minimal subtraction as detailed in Ref.~\cite{ZJ02}, and then set the couplings to their fixed point values. The analytic continuation of divergent loop sums to arbitrary dimension can be found, for example, in Refs.~\cite{EBJZ85,ZJ02,WS16}.

The one-loop result for $h_{k=0}$ can be written
\beq
h(\varphi) = \frac{1}{\sqrt{A}}\left(\frac{1}{2}\pi^2 + \frac{R}{2} \varphi^2 + \frac{U}{4!} \varphi^4 \right),
\eeq
where $R$ and $U$ are dimensionless universal quantities which form a power series in $\epsilon$. These constants will also depend on $\tau$, and our expression for them is in terms of integrals over Riemann theta functions which need to be evaluated numerically. As we will justify below, we need to calculate $R$ to order $\epsilon$ and $U$ to order $\epsilon^2$ to obtain the spectrum to one-loop. Given the commutation relation $[\varphi,\pi] = i$, the momentum acts on the zero-mode functional as
\beq
\pi^2 \Psi[\varphi] = - \frac{d^2}{d \varphi^2} \Psi[\varphi]
\eeq
Therefore, at leading order in the $\epsilon$-expansion, the low-energy spectrum of the Ising model on the torus maps onto the spectrum of a one-dimensional quantum anharmonic oscillator with universal coefficients.

In spite of the effective Hamiltonian being an ordinary series in $\epsilon$, the oscillator is strongly coupled for small $\epsilon$. This can be seen by performing the canonical transformation $\varphi \rightarrow U^{-1/6}\varphi$ and $\pi \rightarrow U^{1/6}\pi$, which takes
\bea
&& \left(  \frac{\pi^2}{2} + \frac{R}{2} \varphi^2 + \frac{U}{4!} \varphi^4 \right) \nn
&& \qquad \qquad \longrightarrow U^{1/3} \left( \frac{\pi^2}{2} + \frac{RU^{-2/3}}{2} \varphi^2 + \frac{1}{4!} \varphi^4 \right)
\eea
Since both $R$ and $U$ are $\mathcal{O}(\epsilon)$, this latter form implies that the energy eigenvalues are an expansion in $\epsilon^{1/3}$, and the coefficients of the expansion are given by a pure quartic oscillator perturbed by a quadratic term. The latter problem has been widely studied in the literature, and the coefficients of this expansion to high order can be found in Ref. \cite{SCZ99}.

The above form for the effective Hamiltonian shows that the $\epsilon$ expansion on the torus results in a reordering of the perturbation expansion, since powers of $\varphi$ effectively carry a factor of $\epsilon^{-1/6}$. This reordering is what justifies our calculating $R$ to order $\epsilon$ and $U$ to order $\epsilon^2$ above. A detailed analysis shows that the one-loop expansion of the energy levels is accurate to order $\epsilon^{4/3}$, since the leading two-loop correction to the energy is of order $\epsilon^{5/3}$. This leading two-loop correction is to the coefficient $R$, and we also need to add terms of the form $p^2 \varphi^4 + \mathrm{c.c.}$ and $\varphi^6$ to the effective Hamiltonian at the same order.

For the finite momentum states, the effective Hamiltonian also takes the form of a strongly-coupled oscillator, but the coefficients will depend on the momentum. In addition, the effective Hamiltonian will couple different Fock states with the same energy and momentum whenever possible, which can lead to a multi-dimensional Hamiltonian which mixes Fock states; for these Hamiltonians the effect of these off-diagonal terms were computed numerically.

Finally, we note that since the anti-periodic sectors in the Ising$^{\ast}$ transition do not have a zero mode, the calculation is straight-forward. The energy levels are a normal expansion in $\epsilon$, and we simply need to compute the $\mathcal{O}(\epsilon)$ correction to the energy using first-order perturbation theory.

\clearpage
\FloatBarrier
\widetext
\section{Complete low-energy spectrum for Ising CFT with modular parameter $\tau=i$}

%%divided by c values

\begin{table*}[h!]
 \centering
  \begin{tabular}{|c||c|c|c|c|c|c||c|}
  \hline
  $\tau=i$ & $\kappa=0$ & $\kappa=1 \;(\times 4)$  & $\kappa= \sqrt{2} \;(\times 4)$ & $\kappa=2 \;(\times 4)$ & $\kappa=\sqrt{5} \;(\times 8)$ & $\kappa=2\sqrt{2} \;(\times 4)$ & Denomination\\ \hline
  & 0 &  &  &  &  &  & 1\\
  & \cellcolor{lightgray} \textcolor{blue}{1.28} &  &  &  &  &  & $\sigma_T$\\
  & 4.71 &  &  &  &  &  & $\varepsilon_T$\\
  &  & \cellcolor{lightgray} \textcolor{blue}{6.79} &  &  &  &  & $\sigma_T+\Delta \kappa$ \\
  & \cellcolor{lightgray} 8.77 &  &  &  &  &  & $\sigma^{\prime}_T$ \\
  &  &  & \cellcolor{lightgray} \textcolor{blue}{9.44} &  &  &  & \\
  &  & 9.52 &  &  &  &  & $\varepsilon_T+\Delta \kappa$ \\
  &  &  & 11.6 &  &  &  & \\
  & 12.9 &  &  &  &  &  & \\
  &  &  &  & \cellcolor{lightgray} \textcolor{blue}{13.15} &  &  & \\
  &  &  &  & 13.5 &  &  & \\
  &  & \cellcolor{lightgray} 13.6 &  &  &  &  & \\
  & 14.1 &  &  &  &  &  & \\
  & 14.4 &  &  &  &  &  & \\
  $\tilde{E}/c$ &  &  & 14.5 &  &  &  & \\
  &  &  &  &  & \cellcolor{lightgray} \textcolor{blue}{14.67}&  & \\
  &  &  &  & 14.9 &  &  & \\
  &  &  &  &  & 15.4 &  & \\
  &  &  & \cellcolor{lightgray} 15.6&  &  &  & \\
  &  & 16.0 &  &  &  &  & \\ 
  &  & 17.3 &  &  &  &  & \\
  &  &  &  & \cellcolor{lightgray} 17.6 &  &  & \\
  & \cellcolor{lightgray} 17.7 &  &  &  &  &  & \\   
  & 17.9 &  &  &  &  &  & \\
  &  &  & \cellcolor{lightgray} 18.4 &  &  &  &  \\
  &  &  &  &  &  & \cellcolor{lightgray} \textcolor{blue}{18.46} & \\
%   & \cellcolor{lightgray} 18.7 &  &  & \cellcolor{lightgray} 18.7 &  &  &  \\
%   &  & \cellcolor{lightgray} 20.1 &  &  &  &  &  \\
  \hline
  \end{tabular}
 \caption{\label{tab:squareoverc}Low-energy spectrum $\tilde{E}/c=(E-E_0)\sqrt{N}/c$ for the Ising QFT with $\tau=i$ obtained from ED/QMC on the square lattice. $c$ denotes the speed of light~(see Tab.~\ref{tab:speedoflight}). Unshaded (shaded) cells are even (odd) under spin-inversion. Blue colored values are obtained from QMC+ED, the other values from ED alone. The degeneracy of the finite-momentum levels is given in brackets, all levels for $\kappa=0$ are not degenerate, some very close levels may, however, be actually degenerate in the thermodynamic limit.
 The given values are obtained by linear fits of the finite-size levels $\tilde{E}_N/c$ as a function of $1/N$ and should be accurate up to variations of the last given digit. Obtaining more precise values is a non-trivial task as the values from QMC are the result of a series of fits and ED data shows larger finite-size effects for higher levels in the spectrum and for larger momentum $\kappa$, where the available finite-size momenta already lie within the non-linear regime of the dispersion relation close to the Brillouin zone boundary.
 The last column shows our denomination of the levels as it was used in the main text. 
 See Tab.~\ref{tab:epsilondataoverc} for a comparison with $\epsilon$-expansion results. 
 }
\end{table*}

%%epsilon-expansion
\begin{table*}[h!]
 \centering
  \begin{tabular}{|c||c|c|c|c|c|c|c|c||c|}
  \hline
  $\tau=i$ & $\kappa=0$ & $\kappa_2=0$ & $\kappa=1$ & $\kappa= \sqrt{2}$ & $\kappa_2= \sqrt{2}$ & $\kappa=2$ & $\kappa=\sqrt{5}$ & $\kappa=2\sqrt{2}$ & Denomination\\ \hline
  & 0 &  &  &  &  &  &  &  & 1\\
  & \cellcolor{lightgray} 1.825 &  &  &  &  &  &  &  & $\sigma_T$\\
  & 5.16 &  &  &  &  &  &  &  & $\varepsilon_T$\\
  &  &  & \cellcolor{lightgray} 6.40 &  &  &  &  &  & $\sigma_T+\Delta \kappa$ \\
  &  &  & 8.88 &  &  &  &  &  & $\varepsilon_T+\Delta \kappa$ \\
  &  &  &  & \cellcolor{lightgray} 9.01 &  &  &  &  & \\
  & \cellcolor{lightgray} 9.02  & &  &  &  &  &  &  & $\sigma^{\prime}_T$ \\
  &  &  &  & 11.35 &  &  &  &  & \\
  &  &  & \cellcolor{lightgray} 12.41 &  &  &  &  &  & \\
  &  &  &  &  &  & \cellcolor{lightgray} 12.60 &  &  & \\
  &  &  &  &  &  & 12.72 &  &  & \\
  $\tilde{E}/c$ &  & 12.95 &  &  & 12.95 &  &  &  & \\
  & 13.39 &  &  &  &  &  &  &  & \\
  &  &  &  &  &  &  & \cellcolor{lightgray} 14.15 &  & \\
  &  &  &  & \cellcolor{lightgray} 14.867 &  &  &  &  & \\
  &  &  &  &  &  & 14.873 &  &  & \\
  &  &  & 15.51 &  &  &  & 15.51 &  & \\
  &  &  &  &  &  & \cellcolor{lightgray} 15.83 &  &  & \\
  &  & \cellcolor{lightgray} 16.00 &  &  & \cellcolor{lightgray} 16.00 &  &  &  & \\
  &  &  &  &  &  &  & 16.32 &  & \\
  &  &  & 16.50 &  &  &  &  &  & \\
  &  &  &  &  &  &  &  & \cellcolor{lightgray} 17.78 & \\
%   &  &  &  &  &  &  &  & 17.93 & \\
%   &  & 18.08 &  &  &  & 18.08 &  &  & \\
%   & \cellcolor{lightgray} 18.18 &  &  &  &  &  &  &  & \\
  \hline
  \end{tabular}
 \caption{\label{tab:epsilondataoverc}Low-energy spectrum for the Ising QFT with modular parameter $\tau=i$ from $\epsilon$-expansion. The notation $\kappa_2$ indicates "two-particle" states (but this distinction loses meaning for higher $\kappa$).}
\end{table*}

%\clearpage
\FloatBarrier
\section{Complete low-energy spectrum for Ising CFT with modular parameter $\tau=\frac{1}{2} + \frac{\sqrt{3}}{2} i$}
%divided by c values

\begin{table*}[h!]
 \centering
  \begin{tabular}{|c||c|c|c|c|c||c|}
  \hline
  $\tau=\frac{1}{2} + \frac{\sqrt{3}}{2} i$ & $\kappa=0$ & $\kappa=1 \;(\times 6)$  & $\kappa= \sqrt{3} \;(\times 6)$ & $\kappa=2 \;(\times 6)$ & $\kappa=\sqrt{7} \;(\times 12)$ & Denomination\\ \hline
  & 0 &  &  &  &  & 1\\
  & \cellcolor{lightgray} \textcolor{blue}{1.35} &  &  &  &  & $\sigma_T$\\
  & 5.03 &  &  &  &   & $\varepsilon_T$\\
  &  & \cellcolor{lightgray} \textcolor{blue}{7.77} &  &  &  & $\sigma_T+\Delta \kappa$ \\
  & \cellcolor{lightgray} 9.33 &  &  &  &  & $\sigma^{\prime}_T$ \\
  &  & 10.53 &  &  &   & $\varepsilon_T+\Delta \kappa$\\
  &  &  & \cellcolor{lightgray} \textcolor{blue}{13.14} &  &  & \\
  & 14.5 &  &  &  &  &  \\
  &  &  & 14.6 &  &   & \\
  &  & \cellcolor{lightgray} 14.8 &  &  &  &\\
  & 15.1 $(\times 2)$ &  &  &  &  &  \\
  &  &  &  & \cellcolor{lightgray} \textcolor{blue}{15.10}&  & \\
  &  &  &  & 15.9 &   & \\
  &  & 16.1 &  &  &   & \\
  &  &  &  & 16.2 &   & \\
  &  &  & 16.7 &  &   & \\
  $\tilde{E}/c$ & 18.3 &  &  &  &  & \\
  &  &  & \cellcolor{lightgray} 18.6 &  &  &\\
  &  &  &  &  & \cellcolor{lightgray} \textcolor{blue}{19.82} & \\
  & \cellcolor{lightgray} 19.8 $(\times 2)$ &  &  &  &  & \\
%   &  &  &  & \cellcolor{lightgray} 20.1 &  & \\
%   &  & 20.2 &  &  &  &  \\
%   &  & \cellcolor{lightgray} 20.2 &  &  &  &\\
%   &  &  & \cellcolor{lightgray} 20.3 &  &  &\\
%   & \cellcolor{lightgray} 20.5 &  &  &  &  & \\
%   &  &  &  & \cellcolor{lightgray} 20.7 &  & \\
%   & \cellcolor{lightgray} 21.3  &  &  &  &  & \\
%   &  & 21.3 &  &  &   & \\
%   &  & 21.8 &  &  &   & \\
%   &  &  & 22.9 &  &   & \\
%   &  &  & 23.0 &  &   & \\
%   &  & \cellcolor{lightgray} 23.2 &  &  &  & \\
%   & \cellcolor{lightgray} 25.0 $(\times 2)$ &  &  &  &  & \\
%   &  &  & \cellcolor{lightgray} 25.2 &  &  & \\
%   & 25.2 $(\times 2)$ &  &  &  &  & \\
  \hline
  \end{tabular}
 \caption{\label{tab:triangularoverc}Low-energy spectrum for the Ising QFT with $\tau=\frac{1}{2} + \frac{\sqrt{3}}{2} i$ obtained from ED/QMC on the triangular lattice. See Tab.~\ref{tab:squareoverc} for further details and Tab.~\ref{tab:epsilondataoverc_hex} for a comparison with $\epsilon$-expansion results.
 }
\end{table*}

\begin{table*}[h!]
 \centering
  \begin{tabular}{|c||c|c|c|c|c||c|}
  \hline
  $\tau=\frac{1}{2} + \frac{\sqrt{3}}{2} i$ & $\kappa=0$ & $\kappa=1 \;(\times 6)$  & $\kappa= \sqrt{3} \;(\times 6)$ & $\kappa=2 \;(\times 6)$ & $\kappa=\sqrt{7} \;(\times 12)$ & Denomination\\ \hline
  & 0 &  &  &  &  & 1\\
  & \cellcolor{lightgray} 1.96 &  &  &  &  & $\sigma_T$\\
  & 5.55 &  &  &  &   & $\varepsilon_T$\\
  &  & \cellcolor{lightgray} 7.38 &  &  &  & $\sigma_T+\Delta \kappa$ \\
  & \cellcolor{lightgray} 9.72 &  &  &  &  & $\sigma^{\prime}_T$ \\
  &  & 10.02 &  &  &   & $\varepsilon_T+\Delta \kappa$\\
  &  &  & \cellcolor{lightgray} 12.68 &  &  & \\
  &  & \cellcolor{lightgray} 13.83 &  &  &  &\\
  & 14.31  &  &  &  &  &  \\
  & 14.44  &  &  &  &  &  \\
  &  &  &  & \cellcolor{lightgray} 14.54 &  & \\
  $\tilde{E}/c$ &  &  &  & 14.67 &   & \\
  &  & 14.90 & 14.90 &  &   & \\
  &  &  & 15.08 &  &   & \\
  & 15.23  &  &  &  &  & \\
  & \cellcolor{lightgray} 16.66 &  &  &  &  & \\
  &  &  &  & 16.96 &  & \\
  & \cellcolor{lightgray} 17.59 &  &  &  &  & \\
  &  &  &  & \cellcolor{lightgray} 17.97 &  & \\
  &  & \cellcolor{lightgray} 18.13 & \cellcolor{lightgray} 18.13 &  &  & \\
  &  & 18.24 &  &  &  & \\
  &  &  & \cellcolor{lightgray} 18.84 &  &  & \\
  &  &  &  &  & \cellcolor{lightgray} 19.28 & \\
  & \cellcolor{lightgray} 19.61 &  &  &  &  & \\
  \hline
  \end{tabular}
 \caption{\label{tab:epsilondataoverc_hex}Low-energy spectrum for the Ising QFT with modular parameter $\tau=\frac{1}{2} + \frac{\sqrt{3}}{2}i$ from $\epsilon$-expansion.}
\end{table*}

%\clearpage
\FloatBarrier
\section{Complete low-energy spectrum for Ising* CFT with $\tau=i$}

\begin{table*}[h!]
{%
\newcommand{\mc}[3]{\multicolumn{#1}{#2}{#3}}
\begin{center}
\begin{tabular}{|c||c|c|c|c|c|c|c|c|c||c|}
 \hline
   & \mc{3}{c|}{(P,P)} & \mc{3}{c|}{(P,A)/(A,P)} & \mc{3}{c||}{(A,A)} & \\
 \hline
 $\tau=i$ & $\kappa=0$ & $\kappa=1$ & $\kappa=\sqrt{2}$ & $\kappa=0$ & $\kappa=1$ & $\kappa=\sqrt{2}$ & $\kappa=0$ & $\kappa=1$ & $\kappa=\sqrt{2}$ & Denomination\\
 \hline
   & 0 &  &  &  &  &  &  &  &  & 1 \\
   &  &  &  & 0.45 &  &  &  &  &  &  $1'_T$ \\
   &  &  &  &  &  &  & 0.62 &  &  &  $1''_T$ \\
   & 4.71 &  &  &  &  &  &  &  &  &  $\varepsilon_T$ \\
   &  &  &  &  & 7.3 &  &  &  & & \\
   &  &  &  & 8.5 &  &  &  &  & & \\
 $\tilde{E}/c$&  &  &  &  &  &  & 9.0 &  &  & \\
   &  & 9.52 &  &  &  &  &  &  &  & \\
   &  &  &  &  &  &  &  &  & 9.6 & \\
   &  &  &  &  &  &  &  & 10.3 &  & \\
   &  &  &  &  & 10.4 &  &  &  &  & \\
   &  &  &  &  &  & 11.4 &  &  &  & \\
   &  &  & 11.6 &  &  &  &  &  &  & \\
 \hline             
\end{tabular}
 \caption{Low-energy spectrum for the Ising* CFT with $\tau=i$ obtained from ED on the square lattice. The four distinct topological sectors are indicated by the corresponding boundary conditions (P,A) etc. in the two directions around the torus, where A(P) denotes (anti-)periodic boundary conditions~(See main text for further details).
 The four lowest-lying levels constitute the topological four-fold degenerate groundstate manifold in the Toric Code phase and are still remarkably low in energy at criticality. See Tab.~\ref{tab:isingstarepsilon} for a comparison with $\epsilon$-expansion results and Tab.~\ref{tab:squareoverc} for further details.} %A discrepancy to the $\varepsilon$-expansion (see data\_tc.pdf) is the second $\kappa=0$ state in the (A,A) sector.}
\end{center}
}%
\end{table*}

\begin{table*}[h!]
{%
\newcommand{\mc}[3]{\multicolumn{#1}{#2}{#3}}
\begin{center}
\begin{tabular}{|c||c|c|c|c|c|c|c|c|c||c|}
 \hline
   & \mc{3}{c|}{(P,P)} & \mc{3}{c|}{(P,A)/(A,P)} & \mc{3}{c||}{(A,A)} &\\
 \hline
 $\tau=i$ & $\kappa=0$ & $\kappa=1$ & $\kappa=\sqrt{2}$ & $\kappa=0$ & $\kappa=1$ & $\kappa=\sqrt{2}$ & $\kappa=0$ & $\kappa=1$ & $\kappa=\sqrt{2}$ & Denomination \\
 \hline
   & 0 &  &  &  &  &  &  &  &  & 1 \\
   &  &  &  & 0.19 &  &  &  &  &  & $1'_T$ \\
   &  &  &  &  &  &  & 0.35 &  &  & $1''_T$ \\
   & 5.16 &  &  &  &  &  &  &  &  & $\varepsilon_T$ \\
   $\tilde{E}/c$ &  &  &  &  & 6.88 &  &  &  &  & \\
   &  &  &  & 7.54 &  &  &  &  &  & \\
   &  & 8.88 &  &  &  &  &  &  &  & \\
   &  &  &  &  &  &  &  &  & 9.18 & \\
   &  &  &  &  &  &  & 9.75 & 9.75 &  & \\
   &  &  & 11.35 &  &  &  &  &  &  & \\
 \hline             
\end{tabular}
 \caption{\label{tab:isingstarepsilon}Low-energy spectrum for the Ising* CFT with $\tau=i$ obtained from $\epsilon$-expansion.} %A discrepancy to the $\varepsilon$-expansion (see data\_tc.pdf) is the second $\kappa=0$ state in the (A,A) sector.}
\end{center}
}%
\end{table*}

\end{document}

%% file: IsingIsingStar.bbl
%merlin.mbs apsrev4-1.bst 2010-07-25 4.21a (PWD, AO, DPC) hacked
%Control: key (0)
%Control: author (0) dotless jnrlst
%Control: editor formatted (1) identically to author
%Control: production of article title (0) allowed
%Control: page (1) range
%Control: year (0) verbatim
%Control: production of eprint (0) enabled
%

%% file: IsingIsingStar_arxiv_Single.bbl
\begin{thebibliography}{65}%
\makeatletter
\providecommand \@ifxundefined [1]{%
 \@ifx{#1\undefined}
}%
\providecommand \@ifnum [1]{%
 \ifnum #1\expandafter \@firstoftwo
 \else \expandafter \@secondoftwo
 \fi
}%
\providecommand \@ifx [1]{%
 \ifx #1\expandafter \@firstoftwo
 \else \expandafter \@secondoftwo
 \fi
}%
\providecommand \natexlab [1]{#1}%
\providecommand \enquote  [1]{``#1''}%
\providecommand \bibnamefont  [1]{#1}%
\providecommand \bibfnamefont [1]{#1}%
\providecommand \citenamefont [1]{#1}%
\providecommand \href@noop [0]{\@secondoftwo}%
\providecommand \href [0]{\begingroup \@sanitize@url \@href}%
\providecommand \@href[1]{\@@startlink{#1}\@@href}%
\providecommand \@@href[1]{\endgroup#1\@@endlink}%
\providecommand \@sanitize@url [0]{\catcode `\\12\catcode `\$12\catcode
  `\&12\catcode `\#12\catcode `\^12\catcode `\_12\catcode `\%12\relax}%
\providecommand \@@startlink[1]{}%
\providecommand \@@endlink[0]{}%
\providecommand \url  [0]{\begingroup\@sanitize@url \@url }%
\providecommand \@url [1]{\endgroup\@href {#1}{\urlprefix }}%
\providecommand \urlprefix  [0]{URL }%
\providecommand \Eprint [0]{\href }%
\providecommand \doibase [0]{http://dx.doi.org/}%
\providecommand \selectlanguage [0]{\@gobble}%
\providecommand \bibinfo  [0]{\@secondoftwo}%
\providecommand \bibfield  [0]{\@secondoftwo}%
\providecommand \translation [1]{[#1]}%
\providecommand \BibitemOpen [0]{}%
\providecommand \bibitemStop [0]{}%
\providecommand \bibitemNoStop [0]{.\EOS\space}%
\providecommand \EOS [0]{\spacefactor3000\relax}%
\providecommand \BibitemShut  [1]{\csname bibitem#1\endcsname}%
\let\auto@bib@innerbib\@empty
%</preamble>
\bibitem [{\citenamefont {Senthil}\ \emph {et~al.}(2004)\citenamefont
  {Senthil}, \citenamefont {Vishwanath}, \citenamefont {Balents}, \citenamefont
  {Sachdev},\ and\ \citenamefont {Fisher}}]{Senthil2004}%
  \BibitemOpen
  \bibfield  {author} {\bibinfo {author} {\bibfnamefont {T.}~\bibnamefont
  {Senthil}}, \bibinfo {author} {\bibfnamefont {A.}~\bibnamefont {Vishwanath}},
  \bibinfo {author} {\bibfnamefont {L.}~\bibnamefont {Balents}}, \bibinfo
  {author} {\bibfnamefont {S.}~\bibnamefont {Sachdev}}, \ and\ \bibinfo
  {author} {\bibfnamefont {M.~P.~A.}\ \bibnamefont {Fisher}},\ }\bibfield
  {title} {\enquote {\bibinfo {title} {{Deconfined Quantum Critical Points}},}\
  }\href {\doibase 10.1126/science.1091806} {\bibfield  {journal} {\bibinfo
  {journal} {Science}\ }\textbf {\bibinfo {volume} {303}},\ \bibinfo {pages}
  {1490--1494} (\bibinfo {year} {2004})}\BibitemShut {NoStop}%
\bibitem [{\citenamefont {Sandvik}(2007)}]{Sandvik2007}%
  \BibitemOpen
  \bibfield  {author} {\bibinfo {author} {\bibfnamefont {A.~W.}\ \bibnamefont
  {Sandvik}},\ }\bibfield  {title} {\enquote {\bibinfo {title} {{Evidence for
  deconfined quantum criticality in a two-dimensional Heisenberg model with
  four-spin interactions}},}\ }\href {\doibase 10.1103/PhysRevLett.98.227202}
  {\bibfield  {journal} {\bibinfo  {journal} {Phys. Rev. Lett.}\ }\textbf
  {\bibinfo {volume} {98}},\ \bibinfo {pages} {227202} (\bibinfo {year}
  {2007})},\ \Eprint {http://arxiv.org/abs/0611343} {arXiv:0611343 [cond-mat]}
  \BibitemShut {NoStop}%
\bibitem [{\citenamefont {El-Showk}\ \emph {et~al.}(2012)\citenamefont
  {El-Showk}, \citenamefont {Paulos}, \citenamefont {Poland}, \citenamefont
  {Rychkov}, \citenamefont {Simmons-Duffin},\ and\ \citenamefont
  {Vichi}}]{El-Showk2012}%
  \BibitemOpen
  \bibfield  {author} {\bibinfo {author} {\bibfnamefont {S.}~\bibnamefont
  {El-Showk}}, \bibinfo {author} {\bibfnamefont {M.~F.}\ \bibnamefont
  {Paulos}}, \bibinfo {author} {\bibfnamefont {D.}~\bibnamefont {Poland}},
  \bibinfo {author} {\bibfnamefont {S.}~\bibnamefont {Rychkov}}, \bibinfo
  {author} {\bibfnamefont {D.}~\bibnamefont {Simmons-Duffin}}, \ and\ \bibinfo
  {author} {\bibfnamefont {A.}~\bibnamefont {Vichi}},\ }\bibfield  {title}
  {\enquote {\bibinfo {title} {{Solving the 3D Ising model with the conformal
  bootstrap}},}\ }\href {\doibase 10.1103/PhysRevD.86.025022} {\bibfield
  {journal} {\bibinfo  {journal} {Phys. Rev. D}\ }\textbf {\bibinfo {volume}
  {86}},\ \bibinfo {pages} {025022} (\bibinfo {year} {2012})},\ \Eprint
  {http://arxiv.org/abs/1403.4545} {arXiv:1403.4545} \BibitemShut {NoStop}%
\bibitem [{\citenamefont {El-Showk}\ \emph {et~al.}(2014)\citenamefont
  {El-Showk}, \citenamefont {Paulos}, \citenamefont {Poland}, \citenamefont
  {Rychkov}, \citenamefont {Simmons-Duffin},\ and\ \citenamefont
  {Vichi}}]{El-showk2014}%
  \BibitemOpen
  \bibfield  {author} {\bibinfo {author} {\bibfnamefont {S.}~\bibnamefont
  {El-Showk}}, \bibinfo {author} {\bibfnamefont {M.~F.}\ \bibnamefont
  {Paulos}}, \bibinfo {author} {\bibfnamefont {D.}~\bibnamefont {Poland}},
  \bibinfo {author} {\bibfnamefont {S.}~\bibnamefont {Rychkov}}, \bibinfo
  {author} {\bibfnamefont {D.}~\bibnamefont {Simmons-Duffin}}, \ and\ \bibinfo
  {author} {\bibfnamefont {A.}~\bibnamefont {Vichi}},\ }\bibfield  {title}
  {\enquote {\bibinfo {title} {{Solving the 3d Ising Model with the Conformal
  Bootstrap II. $c$-Minimization and Precise Critical Exponents}},}\ }\href
  {\doibase 10.1007/s10955-014-1042-7} {\bibfield  {journal} {\bibinfo
  {journal} {J. Stat. Phys.}\ }\textbf {\bibinfo {volume} {157}},\ \bibinfo
  {pages} {869--914} (\bibinfo {year} {2014})},\ \Eprint
  {http://arxiv.org/abs/1403.4545} {arXiv:1403.4545} \BibitemShut {NoStop}%
\bibitem [{\citenamefont {Holzhey}\ \emph {et~al.}(1994)\citenamefont
  {Holzhey}, \citenamefont {Larsen},\ and\ \citenamefont
  {Wilczek}}]{Holzhey1994}%
  \BibitemOpen
  \bibfield  {author} {\bibinfo {author} {\bibfnamefont {C.}~\bibnamefont
  {Holzhey}}, \bibinfo {author} {\bibfnamefont {F.}~\bibnamefont {Larsen}}, \
  and\ \bibinfo {author} {\bibfnamefont {F.}~\bibnamefont {Wilczek}},\
  }\bibfield  {title} {\enquote {\bibinfo {title} {{Geometric and renormalized
  entropy in conformal field theory}},}\ }\href {\doibase
  10.1016/0550-3213(94)90402-2} {\bibfield  {journal} {\bibinfo  {journal}
  {Nucl. Phys. B}\ }\textbf {\bibinfo {volume} {424}},\ \bibinfo {pages}
  {443--467} (\bibinfo {year} {1994})}\BibitemShut {NoStop}%
\bibitem [{\citenamefont {Calabrese}\ and\ \citenamefont
  {Cardy}(2004)}]{Calabrese2004}%
  \BibitemOpen
  \bibfield  {author} {\bibinfo {author} {\bibfnamefont {P.}~\bibnamefont
  {Calabrese}}\ and\ \bibinfo {author} {\bibfnamefont {J.}~\bibnamefont
  {Cardy}},\ }\bibfield  {title} {\enquote {\bibinfo {title} {{Entanglement
  entropy and quantum field theory}},}\ }\href {\doibase
  10.1088/1742-5468/2004/06/P06002} {\bibfield  {journal} {\bibinfo  {journal}
  {J. Stat. Mech. Theor. Exp.}\ }\textbf {\bibinfo {volume} {2004}},\ \bibinfo
  {pages} {P06002} (\bibinfo {year} {2004})}\BibitemShut {NoStop}%
\bibitem [{\citenamefont {L{\"{a}}uchli}(2013)}]{Laeuchli2013}%
  \BibitemOpen
  \bibfield  {author} {\bibinfo {author} {\bibfnamefont {A.~M.}\ \bibnamefont
  {L{\"{a}}uchli}},\ }\bibfield  {title} {\enquote {\bibinfo {title} {{Operator
  content of real-space entanglement spectra at conformal critical points}},}\
  }\href {http://arxiv.org/abs/1303.0741} {\bibfield  {journal} {\bibinfo
  {journal} {ArXiv e-prints}\ } (\bibinfo {year} {2013})},\ \Eprint
  {http://arxiv.org/abs/1303.0741} {arXiv:1303.0741} \BibitemShut {NoStop}%
\bibitem [{\citenamefont {Kallin}\ \emph {et~al.}(2013)\citenamefont {Kallin},
  \citenamefont {Hyatt}, \citenamefont {Singh},\ and\ \citenamefont
  {Melko}}]{Kallin2013}%
  \BibitemOpen
  \bibfield  {author} {\bibinfo {author} {\bibfnamefont {A.~B.}\ \bibnamefont
  {Kallin}}, \bibinfo {author} {\bibfnamefont {K.}~\bibnamefont {Hyatt}},
  \bibinfo {author} {\bibfnamefont {R.~R.~P.}\ \bibnamefont {Singh}}, \ and\
  \bibinfo {author} {\bibfnamefont {R.~G.}\ \bibnamefont {Melko}},\ }\bibfield
  {title} {\enquote {\bibinfo {title} {{Entanglement at a Two-Dimensional
  Quantum Critical Point: A Numerical Linked-Cluster Expansion Study}},}\
  }\href {\doibase 10.1103/PhysRevLett.110.135702} {\bibfield  {journal}
  {\bibinfo  {journal} {Phys. Rev. Lett.}\ }\textbf {\bibinfo {volume} {110}},\
  \bibinfo {pages} {135702} (\bibinfo {year} {2013})}\BibitemShut {NoStop}%
\bibitem [{\citenamefont {Bueno}\ \emph {et~al.}(2015)\citenamefont {Bueno},
  \citenamefont {Myers},\ and\ \citenamefont {Witczak-Krempa}}]{Bueno2015}%
  \BibitemOpen
  \bibfield  {author} {\bibinfo {author} {\bibfnamefont {P.}~\bibnamefont
  {Bueno}}, \bibinfo {author} {\bibfnamefont {R.~C.}\ \bibnamefont {Myers}}, \
  and\ \bibinfo {author} {\bibfnamefont {W.}~\bibnamefont {Witczak-Krempa}},\
  }\bibfield  {title} {\enquote {\bibinfo {title} {{Universality of Corner
  Entanglement in Conformal Field Theories}},}\ }\href {\doibase
  10.1103/PhysRevLett.115.021602} {\bibfield  {journal} {\bibinfo  {journal}
  {Phys. Rev. Lett.}\ }\textbf {\bibinfo {volume} {115}},\ \bibinfo {pages}
  {021602} (\bibinfo {year} {2015})}\BibitemShut {NoStop}%
\bibitem [{Note1()}]{Note1}%
  \BibitemOpen
  \bibinfo {note} {In a corresponding classical statistical mechanics language,
  we are discussing the spectrum of the logarithm of the transfer matrix in the
  limit of an infinitely long square (or hexagonal) rod~(c.f.~left part of
  Fig.~\ref {fig:TFI_geometry}). The transfer matrix acts along the infinite
  rod direction.}\BibitemShut {Stop}%
\bibitem [{\citenamefont {Cardy}(1984)}]{Cardy1984}%
  \BibitemOpen
  \bibfield  {author} {\bibinfo {author} {\bibfnamefont {J.~L.}\ \bibnamefont
  {Cardy}},\ }\bibfield  {title} {\enquote {\bibinfo {title} {{Conformal
  invariance and universality in finite-size scaling}},}\ }\href
  {http://stacks.iop.org/0305-4470/17/i=7/a=003} {\bibfield  {journal}
  {\bibinfo  {journal} {J. Phys. A. Math. Gen.}\ }\textbf {\bibinfo {volume}
  {17}},\ \bibinfo {pages} {L385--L387} (\bibinfo {year} {1984})}\BibitemShut
  {NoStop}%
\bibitem [{\citenamefont {Feiguin}\ \emph {et~al.}(2007)\citenamefont
  {Feiguin}, \citenamefont {Trebst}, \citenamefont {Ludwig}, \citenamefont
  {Troyer}, \citenamefont {Kitaev}, \citenamefont {Wang},\ and\ \citenamefont
  {Freedman}}]{Feiguin2007}%
  \BibitemOpen
  \bibfield  {author} {\bibinfo {author} {\bibfnamefont {A.}~\bibnamefont
  {Feiguin}}, \bibinfo {author} {\bibfnamefont {S.}~\bibnamefont {Trebst}},
  \bibinfo {author} {\bibfnamefont {A.~W.~W.}\ \bibnamefont {Ludwig}}, \bibinfo
  {author} {\bibfnamefont {M.}~\bibnamefont {Troyer}}, \bibinfo {author}
  {\bibfnamefont {A.}~\bibnamefont {Kitaev}}, \bibinfo {author} {\bibfnamefont
  {Z.}~\bibnamefont {Wang}}, \ and\ \bibinfo {author} {\bibfnamefont {M.~H.}\
  \bibnamefont {Freedman}},\ }\bibfield  {title} {\enquote {\bibinfo {title}
  {{Interacting Anyons in Topological Quantum Liquids: The Golden Chain}},}\
  }\href {\doibase 10.1103/PhysRevLett.98.160409} {\bibfield  {journal}
  {\bibinfo  {journal} {Phys. Rev. Lett.}\ }\textbf {\bibinfo {volume} {98}},\
  \bibinfo {pages} {160409} (\bibinfo {year} {2007})}\BibitemShut {NoStop}%
\bibitem [{\citenamefont {Suwa}\ and\ \citenamefont {Todo}(2015)}]{Suwa2015}%
  \BibitemOpen
  \bibfield  {author} {\bibinfo {author} {\bibfnamefont {H.}~\bibnamefont
  {Suwa}}\ and\ \bibinfo {author} {\bibfnamefont {S.}~\bibnamefont {Todo}},\
  }\bibfield  {title} {\enquote {\bibinfo {title} {{Generalized Moment Method
  for Gap Estimation and Quantum Monte Carlo Level Spectroscopy}},}\ }\href
  {\doibase 10.1103/PhysRevLett.115.080601} {\bibfield  {journal} {\bibinfo
  {journal} {Phys. Rev. Lett.}\ }\textbf {\bibinfo {volume} {115}},\ \bibinfo
  {pages} {080601} (\bibinfo {year} {2015})},\ \Eprint
  {http://arxiv.org/abs/1402.0847} {arXiv:1402.0847} \BibitemShut {NoStop}%
\bibitem [{\citenamefont {Cardy}(1985)}]{Cardy1985}%
  \BibitemOpen
  \bibfield  {author} {\bibinfo {author} {\bibfnamefont {J.~L.}\ \bibnamefont
  {Cardy}},\ }\bibfield  {title} {\enquote {\bibinfo {title} {{Universal
  amplitudes in finite-size scaling: generalisation to arbitrary
  dimensionality}},}\ }\href {\doibase 10.1088/0305-4470/18/13/005} {\bibfield
  {journal} {\bibinfo  {journal} {J. Phys. A. Math. Gen.}\ }\textbf {\bibinfo
  {volume} {18}},\ \bibinfo {pages} {L757--L760} (\bibinfo {year}
  {1985})}\BibitemShut {NoStop}%
\bibitem [{\citenamefont {Alcaraz}\ and\ \citenamefont
  {Herrmann}(1987)}]{Alcaraz1987b}%
  \BibitemOpen
  \bibfield  {author} {\bibinfo {author} {\bibfnamefont {F.~C.}\ \bibnamefont
  {Alcaraz}}\ and\ \bibinfo {author} {\bibfnamefont {H.~J.}\ \bibnamefont
  {Herrmann}},\ }\bibfield  {title} {\enquote {\bibinfo {title} {{Numerical
  difficulties in obtaining 3D critical exponents from Platonic solids}},}\
  }\href {\doibase 10.1088/0305-4470/20/16/048} {\bibfield  {journal} {\bibinfo
   {journal} {J. Phys. A. Math. Gen.}\ }\textbf {\bibinfo {volume} {20}},\
  \bibinfo {pages} {5735--5736} (\bibinfo {year} {1987})}\BibitemShut {NoStop}%
\bibitem [{\citenamefont {Weigel}\ and\ \citenamefont
  {Janke}(2000)}]{Weigel2000c}%
  \BibitemOpen
  \bibfield  {author} {\bibinfo {author} {\bibfnamefont {M.}~\bibnamefont
  {Weigel}}\ and\ \bibinfo {author} {\bibfnamefont {W.}~\bibnamefont {Janke}},\
  }\bibfield  {title} {\enquote {\bibinfo {title} {{Universal
  amplitude-exponent relation for the Ising model on sphere-like lattices}},}\
  }\href {\doibase 10.1209/epl/i2000-00377-0} {\bibfield  {journal} {\bibinfo
  {journal} {Europhys. Lett.}\ }\textbf {\bibinfo {volume} {51}},\ \bibinfo
  {pages} {578--583} (\bibinfo {year} {2000})},\ \Eprint
  {http://arxiv.org/abs/0008292} {arXiv:0008292 [cond-mat]} \BibitemShut
  {NoStop}%
\bibitem [{\citenamefont {Deng}\ and\ \citenamefont
  {Bl{\"{o}}te}(2002)}]{Deng2002}%
  \BibitemOpen
  \bibfield  {author} {\bibinfo {author} {\bibfnamefont {Y.}~\bibnamefont
  {Deng}}\ and\ \bibinfo {author} {\bibfnamefont {H.~W.~J.}\ \bibnamefont
  {Bl{\"{o}}te}},\ }\bibfield  {title} {\enquote {\bibinfo {title} {{Conformal
  Invariance of the Ising Model in Three Dimensions}},}\ }\href {\doibase
  10.1103/PhysRevLett.88.190602} {\bibfield  {journal} {\bibinfo  {journal}
  {Phys. Rev. Lett.}\ }\textbf {\bibinfo {volume} {88}},\ \bibinfo {pages}
  {190602} (\bibinfo {year} {2002})}\BibitemShut {NoStop}%
\bibitem [{\citenamefont {Brower}\ \emph {et~al.}(2013)\citenamefont {Brower},
  \citenamefont {Fleming},\ and\ \citenamefont {Neuberger}}]{Brower2013c}%
  \BibitemOpen
  \bibfield  {author} {\bibinfo {author} {\bibfnamefont {R.~C.}\ \bibnamefont
  {Brower}}, \bibinfo {author} {\bibfnamefont {G.~T.}\ \bibnamefont {Fleming}},
  \ and\ \bibinfo {author} {\bibfnamefont {H.}~\bibnamefont {Neuberger}},\
  }\bibfield  {title} {\enquote {\bibinfo {title} {{Lattice radial
  quantization: 3D Ising}},}\ }\href {\doibase 10.1016/j.physletb.2013.03.009}
  {\bibfield  {journal} {\bibinfo  {journal} {Phys. Lett. B}\ }\textbf
  {\bibinfo {volume} {721}},\ \bibinfo {pages} {299--305} (\bibinfo {year}
  {2013})},\ \Eprint {http://arxiv.org/abs/1212.6190v1} {arXiv:1212.6190v1}
  \BibitemShut {NoStop}%
\bibitem [{\citenamefont {Brower}\ \emph {et~al.}(2016)\citenamefont {Brower},
  \citenamefont {Fleming}, \citenamefont {Gasbarro}, \citenamefont {Raben},
  \citenamefont {Tan},\ and\ \citenamefont {Weinberg}}]{Brower2016}%
  \BibitemOpen
  \bibfield  {author} {\bibinfo {author} {\bibfnamefont {R.~C.}\ \bibnamefont
  {Brower}}, \bibinfo {author} {\bibfnamefont {G.}~\bibnamefont {Fleming}},
  \bibinfo {author} {\bibfnamefont {A.}~\bibnamefont {Gasbarro}}, \bibinfo
  {author} {\bibfnamefont {T.}~\bibnamefont {Raben}}, \bibinfo {author}
  {\bibfnamefont {C.-I.}\ \bibnamefont {Tan}}, \ and\ \bibinfo {author}
  {\bibfnamefont {E.}~\bibnamefont {Weinberg}},\ }\bibfield  {title} {\enquote
  {\bibinfo {title} {{Quantum Finite Elements for Lattice Field Theory}},}\
  }\href {http://arxiv.org/abs/1601.01367} {\bibfield  {journal} {\bibinfo
  {journal} {ArXiv e-prints}\ } (\bibinfo {year} {2016})},\ \Eprint
  {http://arxiv.org/abs/1601.01367} {arXiv:1601.01367} \BibitemShut {NoStop}%
\bibitem [{\citenamefont {Henningson}\ and\ \citenamefont
  {Wyllard}(2007)}]{Henningson2007}%
  \BibitemOpen
  \bibfield  {author} {\bibinfo {author} {\bibfnamefont {M.}~\bibnamefont
  {Henningson}}\ and\ \bibinfo {author} {\bibfnamefont {N.}~\bibnamefont
  {Wyllard}},\ }\bibfield  {title} {\enquote {\bibinfo {title} {{Low-energy
  spectrum of N = 4 super-Yang-Mills on T$^3$: flat connections, bound states
  at threshold, and S-duality}},}\ }\href {\doibase
  10.1088/1126-6708/2007/06/001} {\bibfield  {journal} {\bibinfo  {journal} {J.
  High Energ. Phys.}\ }\textbf {\bibinfo {volume} {2007}},\ \bibinfo {pages}
  {001--001} (\bibinfo {year} {2007})},\ \Eprint {http://arxiv.org/abs/0703172}
  {arXiv:0703172 [hep-th]} \BibitemShut {NoStop}%
\bibitem [{\citenamefont {Henningson}\ and\ \citenamefont
  {Ohlsson}(2011)}]{Henningson2011}%
  \BibitemOpen
  \bibfield  {author} {\bibinfo {author} {\bibfnamefont {M.}~\bibnamefont
  {Henningson}}\ and\ \bibinfo {author} {\bibfnamefont {F.}~\bibnamefont
  {Ohlsson}},\ }\bibfield  {title} {\enquote {\bibinfo {title} {{BPS partition
  functions in N = 4 Yang-Mills theory on T 4}},}\ }\href {\doibase
  10.1007/JHEP03(2011)145} {\bibfield  {journal} {\bibinfo  {journal} {J. High
  Energ. Phys.}\ }\textbf {\bibinfo {volume} {2011}},\ \bibinfo {pages} {145}
  (\bibinfo {year} {2011})},\ \Eprint {http://arxiv.org/abs/1101.5331}
  {arXiv:1101.5331} \BibitemShut {NoStop}%
\bibitem [{\citenamefont {Banerjee}\ \emph {et~al.}(2013)\citenamefont
  {Banerjee}, \citenamefont {Hellerman}, \citenamefont {Maltz},\ and\
  \citenamefont {Shenker}}]{Banerjee2013}%
  \BibitemOpen
  \bibfield  {author} {\bibinfo {author} {\bibfnamefont {S.}~\bibnamefont
  {Banerjee}}, \bibinfo {author} {\bibfnamefont {S.}~\bibnamefont {Hellerman}},
  \bibinfo {author} {\bibfnamefont {J.}~\bibnamefont {Maltz}}, \ and\ \bibinfo
  {author} {\bibfnamefont {S.~H.}\ \bibnamefont {Shenker}},\ }\bibfield
  {title} {\enquote {\bibinfo {title} {{Light states in Chern-Simons theory
  coupled to fundamental matter}},}\ }\href {\doibase 10.1007/JHEP03(2013)097}
  {\bibfield  {journal} {\bibinfo  {journal} {J. High Energ. Phys.}\ }\textbf
  {\bibinfo {volume} {2013}},\ \bibinfo {pages} {97} (\bibinfo {year}
  {2013})},\ \Eprint {http://arxiv.org/abs/1207.4195} {arXiv:1207.4195}
  \BibitemShut {NoStop}%
\bibitem [{\citenamefont {P{\'{e}}rez}\ \emph {et~al.}(2013)\citenamefont
  {P{\'{e}}rez}, \citenamefont {Gonz{\'{a}}lez-Arroyo},\ and\ \citenamefont
  {Okawa}}]{Perez2013}%
  \BibitemOpen
  \bibfield  {author} {\bibinfo {author} {\bibfnamefont {M.~G.}\ \bibnamefont
  {P{\'{e}}rez}}, \bibinfo {author} {\bibfnamefont {A.}~\bibnamefont
  {Gonz{\'{a}}lez-Arroyo}}, \ and\ \bibinfo {author} {\bibfnamefont
  {M.}~\bibnamefont {Okawa}},\ }\bibfield  {title} {\enquote {\bibinfo {title}
  {{Spatial volume dependence for 2+1 dimensional SU(N) Yang-Mills theory}},}\
  }\href {\doibase 10.1007/JHEP09(2013)003} {\bibfield  {journal} {\bibinfo
  {journal} {J. High Energ. Phys.}\ }\textbf {\bibinfo {volume} {2013}},\
  \bibinfo {pages} {3} (\bibinfo {year} {2013})},\ \Eprint
  {http://arxiv.org/abs/1307.5254} {arXiv:1307.5254} \BibitemShut {NoStop}%
\bibitem [{\citenamefont {Shaghoulian}(2016)}]{Shaghoulian2016}%
  \BibitemOpen
  \bibfield  {author} {\bibinfo {author} {\bibfnamefont {E.}~\bibnamefont
  {Shaghoulian}},\ }\bibfield  {title} {\enquote {\bibinfo {title} {{Modular
  forms and a generalized Cardy formula in higher dimensions}},}\ }\href
  {\doibase 10.1103/PhysRevD.93.126005} {\bibfield  {journal} {\bibinfo
  {journal} {Phys. Rev. D}\ }\textbf {\bibinfo {volume} {93}},\ \bibinfo
  {pages} {126005} (\bibinfo {year} {2016})},\ \Eprint
  {http://arxiv.org/abs/1508.02728} {arXiv:1508.02728} \BibitemShut {NoStop}%
\bibitem [{\citenamefont {Henkel}(1986)}]{Henkel1986}%
  \BibitemOpen
  \bibfield  {author} {\bibinfo {author} {\bibfnamefont {M.}~\bibnamefont
  {Henkel}},\ }\bibfield  {title} {\enquote {\bibinfo {title} {{Universal
  ratios of scaling amplitudes in the Hamiltonian limit of the 3D Ising
  model}},}\ }\href {\doibase 10.1088/0305-4470/19/5/006} {\bibfield  {journal}
  {\bibinfo  {journal} {J. Phys. A. Math. Gen.}\ }\textbf {\bibinfo {volume}
  {19}},\ \bibinfo {pages} {L247--L249} (\bibinfo {year} {1986})}\BibitemShut
  {NoStop}%
\bibitem [{\citenamefont {Henkel}(1987)}]{Henkel1987}%
  \BibitemOpen
  \bibfield  {author} {\bibinfo {author} {\bibfnamefont {M.}~\bibnamefont
  {Henkel}},\ }\bibfield  {title} {\enquote {\bibinfo {title} {{Finite size
  scaling and universality in the (2+1)D Ising model}},}\ }\href {\doibase
  10.1088/0305-4470/20/12/041} {\bibfield  {journal} {\bibinfo  {journal} {J.
  Phys. A. Math. Gen.}\ }\textbf {\bibinfo {volume} {20}},\ \bibinfo {pages}
  {3969--3981} (\bibinfo {year} {1987})}\BibitemShut {NoStop}%
\bibitem [{\citenamefont {Cardy}(1987)}]{Cardy1987}%
  \BibitemOpen
  \bibfield  {author} {\bibinfo {author} {\bibfnamefont {J.~L.}\ \bibnamefont
  {Cardy}},\ }\bibfield  {title} {\enquote {\bibinfo {title} {{Anisotropic
  corrections to correlation functions in finite-size systems}},}\ }\href
  {\doibase 10.1016/0550-3213(87)90192-1} {\bibfield  {journal} {\bibinfo
  {journal} {Nucl. Phys. B}\ }\textbf {\bibinfo {volume} {290}},\ \bibinfo
  {pages} {355--362} (\bibinfo {year} {1987})}\BibitemShut {NoStop}%
\bibitem [{\citenamefont {Hamer}(1983)}]{Hamer1999}%
  \BibitemOpen
  \bibfield  {author} {\bibinfo {author} {\bibfnamefont {C.~J.}\ \bibnamefont
  {Hamer}},\ }\bibfield  {title} {\enquote {\bibinfo {title} {{Finite-size
  scaling in the (2+1)D Ising model}},}\ }\href {\doibase
  10.1088/0305-4470/16/6/020} {\bibfield  {journal} {\bibinfo  {journal} {J.
  Phys. A. Math. Gen.}\ }\textbf {\bibinfo {volume} {16}},\ \bibinfo {pages}
  {1257--1266} (\bibinfo {year} {1983})}\BibitemShut {NoStop}%
\bibitem [{\citenamefont {Hamer}(1986)}]{Hamer1986}%
  \BibitemOpen
  \bibfield  {author} {\bibinfo {author} {\bibfnamefont {C.~J.}\ \bibnamefont
  {Hamer}},\ }\bibfield  {title} {\enquote {\bibinfo {title} {{The (2+1)D Ising
  model on a triangular lattice}},}\ }\href {\doibase
  10.1088/0305-4470/19/3/023} {\bibfield  {journal} {\bibinfo  {journal} {J.
  Phys. A. Math. Gen.}\ }\textbf {\bibinfo {volume} {19}},\ \bibinfo {pages}
  {423--435} (\bibinfo {year} {1986})}\BibitemShut {NoStop}%
\bibitem [{\citenamefont {Hamer}(2000)}]{Hamer2000}%
  \BibitemOpen
  \bibfield  {author} {\bibinfo {author} {\bibfnamefont {C.~J.}\ \bibnamefont
  {Hamer}},\ }\bibfield  {title} {\enquote {\bibinfo {title} {{Finite-size
  scaling in the transverse Ising model on a square lattice}},}\ }\href
  {\doibase 10.1088/0305-4470/33/38/303} {\bibfield  {journal} {\bibinfo
  {journal} {J. Phys. A. Math. Gen.}\ }\textbf {\bibinfo {volume} {33}},\
  \bibinfo {pages} {6683--6698} (\bibinfo {year} {2000})},\ \Eprint
  {http://arxiv.org/abs/0007063} {arXiv:0007063 [cond-mat]} \BibitemShut
  {NoStop}%
\bibitem [{\citenamefont {Nishiyama}(2008)}]{Nishiyama2008}%
  \BibitemOpen
  \bibfield  {author} {\bibinfo {author} {\bibfnamefont {Y.}~\bibnamefont
  {Nishiyama}},\ }\bibfield  {title} {\enquote {\bibinfo {title} {{Bound-state
  energy of the three-dimensional Ising model in the broken-symmetry phase:
  Suppressed finite-size corrections}},}\ }\href {\doibase
  10.1103/PhysRevE.77.051112} {\bibfield  {journal} {\bibinfo  {journal} {Phys.
  Rev. E}\ }\textbf {\bibinfo {volume} {77}},\ \bibinfo {pages} {051112}
  (\bibinfo {year} {2008})},\ \Eprint {http://arxiv.org/abs/0804.1586}
  {arXiv:0804.1586} \BibitemShut {NoStop}%
\bibitem [{\citenamefont {Dusuel}\ \emph {et~al.}(2010)\citenamefont {Dusuel},
  \citenamefont {Kamfor}, \citenamefont {Schmidt}, \citenamefont {Thomale},\
  and\ \citenamefont {Vidal}}]{Dusuel2010}%
  \BibitemOpen
  \bibfield  {author} {\bibinfo {author} {\bibfnamefont {S.}~\bibnamefont
  {Dusuel}}, \bibinfo {author} {\bibfnamefont {M.}~\bibnamefont {Kamfor}},
  \bibinfo {author} {\bibfnamefont {K.~P.}\ \bibnamefont {Schmidt}}, \bibinfo
  {author} {\bibfnamefont {R.}~\bibnamefont {Thomale}}, \ and\ \bibinfo
  {author} {\bibfnamefont {J.}~\bibnamefont {Vidal}},\ }\bibfield  {title}
  {\enquote {\bibinfo {title} {{Bound states in two-dimensional spin systems
  near the Ising limit: A quantum finite-lattice study}},}\ }\href {\doibase
  10.1103/PhysRevB.81.064412} {\bibfield  {journal} {\bibinfo  {journal} {Phys.
  Rev. B}\ }\textbf {\bibinfo {volume} {81}},\ \bibinfo {pages} {064412}
  (\bibinfo {year} {2010})},\ \Eprint {http://arxiv.org/abs/0912.1463}
  {arXiv:0912.1463} \BibitemShut {NoStop}%
\bibitem [{\citenamefont {Jalabert}\ and\ \citenamefont
  {Sachdev}(1991)}]{Jalabert1991}%
  \BibitemOpen
  \bibfield  {author} {\bibinfo {author} {\bibfnamefont {R.~A.}\ \bibnamefont
  {Jalabert}}\ and\ \bibinfo {author} {\bibfnamefont {S.}~\bibnamefont
  {Sachdev}},\ }\bibfield  {title} {\enquote {\bibinfo {title} {{Spontaneous
  alignment of frustrated bonds in an anisotropic, three-dimensional Ising
  model}},}\ }\href {\doibase 10.1103/PhysRevB.44.686} {\bibfield  {journal}
  {\bibinfo  {journal} {Phys. Rev. B}\ }\textbf {\bibinfo {volume} {44}},\
  \bibinfo {pages} {686--690} (\bibinfo {year} {1991})}\BibitemShut {NoStop}%
\bibitem [{\citenamefont {Sachdev}\ and\ \citenamefont {Vojta}(1999)}]{SSMV99}%
  \BibitemOpen
  \bibfield  {author} {\bibinfo {author} {\bibfnamefont {S.}~\bibnamefont
  {Sachdev}}\ and\ \bibinfo {author} {\bibfnamefont {M.}~\bibnamefont
  {Vojta}},\ }\bibfield  {title} {\enquote {\bibinfo {title} {{Translational
  symmetry breaking in two-dimensional antiferromagnets and
  superconductors}},}\ }\href {http://arxiv.org/abs/cond-mat/9910231}
  {\bibfield  {journal} {\bibinfo  {journal} {J. Phys. Soc. Jpn.}\ }\textbf
  {\bibinfo {volume} {69}},\ \bibinfo {pages} {Supp. B, 1} (\bibinfo {year}
  {1999})},\ \Eprint {http://arxiv.org/abs/9910231} {arXiv:9910231 [cond-mat]}
  \BibitemShut {NoStop}%
\bibitem [{\citenamefont {Senthil}\ and\ \citenamefont
  {Fisher}(2000)}]{Senthil2000}%
  \BibitemOpen
  \bibfield  {author} {\bibinfo {author} {\bibfnamefont {T.}~\bibnamefont
  {Senthil}}\ and\ \bibinfo {author} {\bibfnamefont {M.~P.~A.}\ \bibnamefont
  {Fisher}},\ }\bibfield  {title} {\enquote {\bibinfo {title} {{$Z_2$ gauge
  theory of electron fractionalization in strongly correlated systems}},}\
  }\href {\doibase 10.1103/PhysRevB.62.7850} {\bibfield  {journal} {\bibinfo
  {journal} {Phys. Rev. B}\ }\textbf {\bibinfo {volume} {62}},\ \bibinfo
  {pages} {7850--7881} (\bibinfo {year} {2000})}\BibitemShut {NoStop}%
\bibitem [{Note2()}]{Note2}%
  \BibitemOpen
  \bibinfo {note} {See Supplemental Material for a definition of the
  Archimedian lattices.}\BibitemShut {Stop}%
\bibitem [{\citenamefont {Bl{\"{o}}te}\ and\ \citenamefont
  {Deng}(2002)}]{Bloete2002}%
  \BibitemOpen
  \bibfield  {author} {\bibinfo {author} {\bibfnamefont {H.~W.~J.}\
  \bibnamefont {Bl{\"{o}}te}}\ and\ \bibinfo {author} {\bibfnamefont
  {Y.}~\bibnamefont {Deng}},\ }\bibfield  {title} {\enquote {\bibinfo {title}
  {{Cluster Monte Carlo simulation of the transverse Ising model}},}\ }\href
  {\doibase 10.1103/PhysRevE.66.066110} {\bibfield  {journal} {\bibinfo
  {journal} {Phys. Rev. E}\ }\textbf {\bibinfo {volume} {66}},\ \bibinfo
  {pages} {066110} (\bibinfo {year} {2002})}\BibitemShut {NoStop}%
\bibitem [{Note3()}]{Note3}%
  \BibitemOpen
  \bibinfo {note} {We have computed the critical point for the Square-Octagon
  lattice as $(h/J)_c = 2.087(7)$ using a continous-time QMC algorithm similar
  to that of~\cite {Bloete2002}.}\BibitemShut {Stop}%
\bibitem [{Note4()}]{Note4}%
  \BibitemOpen
  \bibinfo {note} {See Supplemental Material for a motivation of this $1/N$
  finite-size extrapolation approach.}\BibitemShut {Stop}%
\bibitem [{Note5()}]{Note5}%
  \BibitemOpen
  \bibinfo {note} {See Supplemental Material for further details about the used
  gap estimation procedure for QMC.}\BibitemShut {Stop}%
\bibitem [{\citenamefont {L{\"u}scher}(1982)}]{L82}%
  \BibitemOpen
  \bibfield  {author} {\bibinfo {author} {\bibfnamefont {M.}~\bibnamefont
  {L{\"u}scher}},\ }\bibfield  {title} {\enquote {\bibinfo {title} {A new
  method to compute the spectrum of low-lying states in massless asymptotically
  free field theories},}\ }\href {\doibase
  http://dx.doi.org/10.1016/0370-2693(82)90210-6} {\bibfield  {journal}
  {\bibinfo  {journal} {Phys. Lett. B}\ }\textbf {\bibinfo {volume} {118}},\
  \bibinfo {pages} {391 -- 394} (\bibinfo {year} {1982})}\BibitemShut {NoStop}%
\bibitem [{\citenamefont {Br\'ezin}\ and\ \citenamefont
  {Zinn-Justin}(1985)}]{EBJZ85}%
  \BibitemOpen
  \bibfield  {author} {\bibinfo {author} {\bibfnamefont {E.}~\bibnamefont
  {Br\'ezin}}\ and\ \bibinfo {author} {\bibfnamefont {J.}~\bibnamefont
  {Zinn-Justin}},\ }\bibfield  {title} {\enquote {\bibinfo {title} {Finite size
  effects in phase transitions},}\ }\href {\doibase
  http://dx.doi.org/10.1016/0550-3213(85)90379-7} {\bibfield  {journal}
  {\bibinfo  {journal} {Nucl. Phys. B}\ }\textbf {\bibinfo {volume} {257}},\
  \bibinfo {pages} {867 -- 893} (\bibinfo {year} {1985})}\BibitemShut {NoStop}%
\bibitem [{\citenamefont {Rudnick}\ \emph {et~al.}(1985)\citenamefont
  {Rudnick}, \citenamefont {Guo},\ and\ \citenamefont {Jasnow}}]{RGJ85}%
  \BibitemOpen
  \bibfield  {author} {\bibinfo {author} {\bibfnamefont {J.}~\bibnamefont
  {Rudnick}}, \bibinfo {author} {\bibfnamefont {H.}~\bibnamefont {Guo}}, \ and\
  \bibinfo {author} {\bibfnamefont {D.}~\bibnamefont {Jasnow}},\ }\bibfield
  {title} {\enquote {\bibinfo {title} {Finite-size scaling and the
  renormalization group},}\ }\href {\doibase 10.1007/BF01009013} {\bibfield
  {journal} {\bibinfo  {journal} {J. Stat. Phys.}\ }\textbf {\bibinfo {volume}
  {41}},\ \bibinfo {pages} {353--373} (\bibinfo {year} {1985})}\BibitemShut
  {NoStop}%
\bibitem [{\citenamefont {Bloch}(1958)}]{B58}%
  \BibitemOpen
  \bibfield  {author} {\bibinfo {author} {\bibfnamefont {C.}~\bibnamefont
  {Bloch}},\ }\bibfield  {title} {\enquote {\bibinfo {title} {Sur la th\'eorie
  des perturbations des \'etats li\'es},}\ }\href {\doibase
  http://dx.doi.org/10.1016/0029-5582(58)90116-0} {\bibfield  {journal}
  {\bibinfo  {journal} {Nucl. Phys.}\ }\textbf {\bibinfo {volume} {6}},\
  \bibinfo {pages} {329 -- 347} (\bibinfo {year} {1958})}\BibitemShut {NoStop}%
\bibitem [{\citenamefont {Sk\'ala}\ \emph {et~al.}(1999)\citenamefont
  {Sk\'ala}, \citenamefont {C\'izek},\ and\ \citenamefont {Zamastil}}]{SCZ99}%
  \BibitemOpen
  \bibfield  {author} {\bibinfo {author} {\bibfnamefont {L.}~\bibnamefont
  {Sk\'ala}}, \bibinfo {author} {\bibfnamefont {J.}~\bibnamefont {C\'izek}}, \
  and\ \bibinfo {author} {\bibfnamefont {J.}~\bibnamefont {Zamastil}},\
  }\bibfield  {title} {\enquote {\bibinfo {title} {{Strong coupling
  perturbation expansions for anharmonic oscillators. Numerical results}},}\
  }\href {http://stacks.iop.org/0305-4470/32/i=30/a=314} {\bibfield  {journal}
  {\bibinfo  {journal} {J. Phys. A. Math. Gen.}\ }\textbf {\bibinfo {volume}
  {32}},\ \bibinfo {pages} {5715} (\bibinfo {year} {1999})}\BibitemShut
  {NoStop}%
\bibitem [{Note6()}]{Note6}%
  \BibitemOpen
  \bibinfo {note} {See Supplemental Material, which includes Refs.~\cite
  {B58,K74,ZJ02,EBJZ85,WS16}}\BibitemShut {NoStop}%
\bibitem [{Note7()}]{Note7}%
  \BibitemOpen
  \bibinfo {note} {See Supplemental Material for a listing of the complete
  low-energy torus spectra for the Ising transition from numerics and $\epsilon
  $-expansion.}\BibitemShut {Stop}%
\bibitem [{\citenamefont {Sen}\ \emph {et~al.}(2015)\citenamefont {Sen},
  \citenamefont {Suwa},\ and\ \citenamefont {Sandvik}}]{Sen2015}%
  \BibitemOpen
  \bibfield  {author} {\bibinfo {author} {\bibfnamefont {A.}~\bibnamefont
  {Sen}}, \bibinfo {author} {\bibfnamefont {H.}~\bibnamefont {Suwa}}, \ and\
  \bibinfo {author} {\bibfnamefont {A.~W.}\ \bibnamefont {Sandvik}},\
  }\bibfield  {title} {\enquote {\bibinfo {title} {{Velocity of excitations in
  ordered, disordered, and critical antiferromagnets}},}\ }\href {\doibase
  10.1103/PhysRevB.92.195145} {\bibfield  {journal} {\bibinfo  {journal} {Phys.
  Rev. B}\ }\textbf {\bibinfo {volume} {92}},\ \bibinfo {pages} {195145}
  (\bibinfo {year} {2015})},\ \Eprint {http://arxiv.org/abs/1505.02535}
  {arXiv:1505.02535} \BibitemShut {NoStop}%
\bibitem [{Note8()}]{Note8}%
  \BibitemOpen
  \bibinfo {note} {See Supplemental Material, which includes Refs.~\cite
  {Sen2015,Hamer2000} for the details on the determination of $c$.}\BibitemShut
  {Stop}%
\bibitem [{Note9()}]{Note9}%
  \BibitemOpen
  \bibinfo {note} {For further studies it is worth noticing that $\epsilon
  $-expansion tends to overestimate the $\kappa =0$ levels while levels with
  $\kappa >0$ are commonly underestimated.}\BibitemShut {Stop}%
\bibitem [{\citenamefont {Trebst}\ \emph {et~al.}(2007)\citenamefont {Trebst},
  \citenamefont {Werner}, \citenamefont {Troyer}, \citenamefont {Shtengel},\
  and\ \citenamefont {Nayak}}]{Trebst2007}%
  \BibitemOpen
  \bibfield  {author} {\bibinfo {author} {\bibfnamefont {S.}~\bibnamefont
  {Trebst}}, \bibinfo {author} {\bibfnamefont {P.}~\bibnamefont {Werner}},
  \bibinfo {author} {\bibfnamefont {M.}~\bibnamefont {Troyer}}, \bibinfo
  {author} {\bibfnamefont {K.}~\bibnamefont {Shtengel}}, \ and\ \bibinfo
  {author} {\bibfnamefont {C.}~\bibnamefont {Nayak}},\ }\bibfield  {title}
  {\enquote {\bibinfo {title} {{Breakdown of a Topological Phase: Quantum Phase
  Transition in a Loop Gas Model with Tension}},}\ }\href {\doibase
  10.1103/PhysRevLett.98.070602} {\bibfield  {journal} {\bibinfo  {journal}
  {Phys. Rev. Lett.}\ }\textbf {\bibinfo {volume} {98}},\ \bibinfo {pages}
  {070602} (\bibinfo {year} {2007})},\ \Eprint {http://arxiv.org/abs/0609048}
  {arXiv:0609048 [cond-mat]} \BibitemShut {NoStop}%
\bibitem [{\citenamefont {Vidal}\ \emph {et~al.}(2009)\citenamefont {Vidal},
  \citenamefont {Dusuel},\ and\ \citenamefont {Schmidt}}]{Vidal2009}%
  \BibitemOpen
  \bibfield  {author} {\bibinfo {author} {\bibfnamefont {J.}~\bibnamefont
  {Vidal}}, \bibinfo {author} {\bibfnamefont {S.}~\bibnamefont {Dusuel}}, \
  and\ \bibinfo {author} {\bibfnamefont {K.~P.}\ \bibnamefont {Schmidt}},\
  }\bibfield  {title} {\enquote {\bibinfo {title} {{Low-energy effective theory
  of the toric code model in a parallel magnetic field}},}\ }\href {\doibase
  10.1103/PhysRevB.79.033109} {\bibfield  {journal} {\bibinfo  {journal} {Phys.
  Rev. B}\ }\textbf {\bibinfo {volume} {79}},\ \bibinfo {pages} {033109}
  (\bibinfo {year} {2009})}\BibitemShut {NoStop}%
\bibitem [{\citenamefont {Tupitsyn}\ \emph {et~al.}(2010)\citenamefont
  {Tupitsyn}, \citenamefont {Kitaev}, \citenamefont {Prokof'ev},\ and\
  \citenamefont {Stamp}}]{Tupitsyn2010}%
  \BibitemOpen
  \bibfield  {author} {\bibinfo {author} {\bibfnamefont {I.~S.}\ \bibnamefont
  {Tupitsyn}}, \bibinfo {author} {\bibfnamefont {A.}~\bibnamefont {Kitaev}},
  \bibinfo {author} {\bibfnamefont {N.~V.}\ \bibnamefont {Prokof'ev}}, \ and\
  \bibinfo {author} {\bibfnamefont {P.~C.~E.}\ \bibnamefont {Stamp}},\
  }\bibfield  {title} {\enquote {\bibinfo {title} {{Topological multicritical
  point in the phase diagram of the toric code model and three-dimensional
  lattice gauge Higgs model}},}\ }\href {\doibase 10.1103/PhysRevB.82.085114}
  {\bibfield  {journal} {\bibinfo  {journal} {Phys. Rev. B}\ }\textbf {\bibinfo
  {volume} {82}},\ \bibinfo {pages} {085114} (\bibinfo {year} {2010})},\
  \Eprint {http://arxiv.org/abs/0804.3175v1} {arXiv:0804.3175v1} \BibitemShut
  {NoStop}%
\bibitem [{\citenamefont {Dusuel}\ \emph {et~al.}(2011)\citenamefont {Dusuel},
  \citenamefont {Kamfor}, \citenamefont {Or{\'{u}}s}, \citenamefont {Schmidt},\
  and\ \citenamefont {Vidal}}]{Dusuel2011}%
  \BibitemOpen
  \bibfield  {author} {\bibinfo {author} {\bibfnamefont {S.}~\bibnamefont
  {Dusuel}}, \bibinfo {author} {\bibfnamefont {M.}~\bibnamefont {Kamfor}},
  \bibinfo {author} {\bibfnamefont {R.}~\bibnamefont {Or{\'{u}}s}}, \bibinfo
  {author} {\bibfnamefont {K.~P.}\ \bibnamefont {Schmidt}}, \ and\ \bibinfo
  {author} {\bibfnamefont {J.}~\bibnamefont {Vidal}},\ }\bibfield  {title}
  {\enquote {\bibinfo {title} {{Robustness of a Perturbed Topological
  Phase}},}\ }\href {\doibase 10.1103/PhysRevLett.106.107203} {\bibfield
  {journal} {\bibinfo  {journal} {Phys. Rev. Lett.}\ }\textbf {\bibinfo
  {volume} {106}},\ \bibinfo {pages} {107203} (\bibinfo {year} {2011})},\
  \Eprint {http://arxiv.org/abs/1012.1740} {arXiv:1012.1740} \BibitemShut
  {NoStop}%
\bibitem [{\citenamefont {Wu}\ \emph {et~al.}(2012)\citenamefont {Wu},
  \citenamefont {Deng},\ and\ \citenamefont {Prokof'ev}}]{Wu2012}%
  \BibitemOpen
  \bibfield  {author} {\bibinfo {author} {\bibfnamefont {F.}~\bibnamefont
  {Wu}}, \bibinfo {author} {\bibfnamefont {Y.}~\bibnamefont {Deng}}, \ and\
  \bibinfo {author} {\bibfnamefont {N.}~\bibnamefont {Prokof'ev}},\ }\bibfield
  {title} {\enquote {\bibinfo {title} {{Phase diagram of the toric code model
  in a parallel magnetic field}},}\ }\href {\doibase
  10.1103/PhysRevB.85.195104} {\bibfield  {journal} {\bibinfo  {journal} {Phys.
  Rev. B}\ }\textbf {\bibinfo {volume} {85}},\ \bibinfo {pages} {195104}
  (\bibinfo {year} {2012})}\BibitemShut {NoStop}%
\bibitem [{\citenamefont {Kitaev}(2003)}]{Kitaev2003}%
  \BibitemOpen
  \bibfield  {author} {\bibinfo {author} {\bibfnamefont {A.~Y.}\ \bibnamefont
  {Kitaev}},\ }\bibfield  {title} {\enquote {\bibinfo {title} {{Fault-tolerant
  quantum computation by anyons}},}\ }\href {\doibase
  10.1016/S0003-4916(02)00018-0} {\bibfield  {journal} {\bibinfo  {journal}
  {Ann. Phys.}\ }\textbf {\bibinfo {volume} {303}},\ \bibinfo {pages} {2--30}
  (\bibinfo {year} {2003})}\BibitemShut {NoStop}%
\bibitem [{\citenamefont {Hamma}\ and\ \citenamefont
  {Lidar}(2008)}]{Hamma2008}%
  \BibitemOpen
  \bibfield  {author} {\bibinfo {author} {\bibfnamefont {A.}~\bibnamefont
  {Hamma}}\ and\ \bibinfo {author} {\bibfnamefont {D.~A.}\ \bibnamefont
  {Lidar}},\ }\bibfield  {title} {\enquote {\bibinfo {title} {{Adiabatic
  Preparation of Topological Order}},}\ }\href {\doibase
  10.1103/PhysRevLett.100.030502} {\bibfield  {journal} {\bibinfo  {journal}
  {Phys. Rev. Lett.}\ }\textbf {\bibinfo {volume} {100}},\ \bibinfo {pages}
  {030502} (\bibinfo {year} {2008})},\ \Eprint {http://arxiv.org/abs/0607145}
  {arXiv:0607145 [quant-ph]} \BibitemShut {NoStop}%
\bibitem [{\citenamefont {Carr}(2010)}]{Carr2010}%
  \BibitemOpen
  \bibinfo {editor} {\bibfnamefont {L.}~\bibnamefont {Carr}},\ ed.,\ \href
  {\doibase 10.1201/b10273} {\emph {\bibinfo {title} {{Understanding Quantum
  Phase Transitions}}}},\ \bibinfo {series} {Condensed Matter Physics}, Vol.\
  \bibinfo {volume} {20103812}\ (\bibinfo  {publisher} {CRC Press},\ \bibinfo
  {year} {2010})\BibitemShut {NoStop}%
\bibitem [{Note10()}]{Note10}%
  \BibitemOpen
  \bibinfo {note} {See Supplemental Material for a detailed discussion of the
  mapping}\BibitemShut {NoStop}%
\bibitem [{Note11()}]{Note11}%
  \BibitemOpen
  \bibinfo {note} {See Supplemental Material for a listing of the complete
  low-energy torus spectra for the Ising* transition from numerics and
  $\epsilon $-expansion.}\BibitemShut {Stop}%
\bibitem [{\citenamefont {Wang}\ \emph {et~al.}(2014)\citenamefont {Wang},
  \citenamefont {Corboz},\ and\ \citenamefont {Troyer}}]{Wang2014}%
  \BibitemOpen
  \bibfield  {author} {\bibinfo {author} {\bibfnamefont {L.}~\bibnamefont
  {Wang}}, \bibinfo {author} {\bibfnamefont {P.}~\bibnamefont {Corboz}}, \ and\
  \bibinfo {author} {\bibfnamefont {M.}~\bibnamefont {Troyer}},\ }\bibfield
  {title} {\enquote {\bibinfo {title} {{Fermionic quantum critical point of
  spinless fermions on a honeycomb lattice}},}\ }\href {\doibase
  10.1088/1367-2630/16/10/103008} {\bibfield  {journal} {\bibinfo  {journal}
  {New J. Phys.}\ }\textbf {\bibinfo {volume} {16}},\ \bibinfo {pages} {103008}
  (\bibinfo {year} {2014})}\BibitemShut {NoStop}%
\bibitem [{\citenamefont {Li}\ \emph {et~al.}(2015)\citenamefont {Li},
  \citenamefont {Jiang},\ and\ \citenamefont {Yao}}]{Li2015}%
  \BibitemOpen
  \bibfield  {author} {\bibinfo {author} {\bibfnamefont {Z.-X.}\ \bibnamefont
  {Li}}, \bibinfo {author} {\bibfnamefont {Y.-F.}\ \bibnamefont {Jiang}}, \
  and\ \bibinfo {author} {\bibfnamefont {H.}~\bibnamefont {Yao}},\ }\bibfield
  {title} {\enquote {\bibinfo {title} {{Fermion-sign-free
  Majarana-quantum-Monte-Carlo studies of quantum critical phenomena of Dirac
  fermions in two dimensions}},}\ }\href {\doibase
  10.1088/1367-2630/17/8/085003} {\bibfield  {journal} {\bibinfo  {journal}
  {New J. Phys.}\ }\textbf {\bibinfo {volume} {17}},\ \bibinfo {pages} {085003}
  (\bibinfo {year} {2015})}\BibitemShut {NoStop}%
\bibitem [{\citenamefont {{Klein}}(1974)}]{K74}%
  \BibitemOpen
  \bibfield  {author} {\bibinfo {author} {\bibfnamefont {D.~J.}\ \bibnamefont
  {{Klein}}},\ }\bibfield  {title} {\enquote {\bibinfo {title} {{Degenerate
  perturbation theory}},}\ }\href {\doibase 10.1063/1.1682018} {\bibfield
  {journal} {\bibinfo  {journal} {J. Chem. Phys.}\ }\textbf {\bibinfo {volume}
  {61}},\ \bibinfo {pages} {786--798} (\bibinfo {year} {1974})}\BibitemShut
  {NoStop}%
\bibitem [{\citenamefont {Zinn-Justin}(2002)}]{ZJ02}%
  \BibitemOpen
  \bibfield  {author} {\bibinfo {author} {\bibfnamefont {J.}~\bibnamefont
  {Zinn-Justin}},\ }\href@noop {} {\emph {\bibinfo {title} {Quantum Field
  Theory and Critical Phenomena}}},\ International series of monographs on
  physics\ (\bibinfo  {publisher} {Clarendon Press},\ \bibinfo {year}
  {2002})\BibitemShut {NoStop}%
\bibitem [{\citenamefont {{Whitsitt}}\ and\ \citenamefont
  {{Sachdev}}(2016)}]{WS16}%
  \BibitemOpen
  \bibfield  {author} {\bibinfo {author} {\bibfnamefont {S.}~\bibnamefont
  {{Whitsitt}}}\ and\ \bibinfo {author} {\bibfnamefont {S.}~\bibnamefont
  {{Sachdev}}},\ }\bibfield  {title} {\enquote {\bibinfo {title} {{Transition
  from the $\mathbb{Z}_2$ spin liquid to antiferromagnetic order: spectrum on
  the torus}},}\ }\href {\doibase 10.1103/PhysRevB.94.085134} {\bibfield
  {journal} {\bibinfo  {journal} {Phys. Rev. B}\ }\textbf {\bibinfo {volume}
  {94}},\ \bibinfo {pages} {085134} (\bibinfo {year} {2016})}\BibitemShut
  {NoStop}%
\end{thebibliography}
